\documentclass[twocolumn,trackchanges]{aastex63}

\usepackage{amsmath,amstext}
\usepackage[figure,figure*]{hypcap}
\usepackage{tablefootnote}
\usepackage{newtxmath} 
\usepackage{float}
\usepackage{xcolor}

\shorttitle{Photometric Metallicity and Carbon Abundance Distributions with S-PLUS}
\shortauthors{Whitten et al.}

\begin{document}

\title{The Photometric Metallicity and Carbon Distributions of the Milky Way’s Halo and Solar Neighborhood from S-PLUS Observations of SDSS Stripe 82}

\author[0000-0002-9594-6143]{Devin D.\ Whitten}
\affiliation{Department of Physics and JINA Center for the Evolution of the Elements, University of Notre Dame, Notre Dame, IN 46556, USA}

\author[0000-0003-4479-1265]{Vinicius M.\ Placco}
\affiliation{Community Science and Data Center/NSF’s NOIRLab, 
             950 N. Cherry Ave., Tucson, AZ 85719, USA}
\affiliation{Department of Physics and JINA Center for the Evolution of the Elements, University of Notre Dame, Notre Dame, IN 46556, USA}

\author[0000-0003-4573-6233]{Timothy C.\ Beers}
\affiliation{Department of Physics and JINA Center for the Evolution of the Elements, University of Notre Dame, Notre Dame, IN 46556, USA}

\author[0000-0002-8072-7511]{Deokkeun An}
\affiliation{Department of Science Education, Ewha Womans University, 52 Ewhayeodae-gil, Seodaemun-gu, Seoul 03760, South Korea}

\author{Young Sun Lee}
\affiliation{Department of Astronomy and Space Science, Chungnam
National University, Daejeon 34134, South Korea}

\author{Felipe~Almeida-Fernandes}
\affiliation{Departamento de Astronomia, Instituto de Astronomia, Geof\'isica e Ci\^encias Atmosf\'ericas da USP, Cidade \\ Universit\'aria, 05508-900, S\~ao Paulo, SP, Brazil}

\author[0000-0001-7907-7884]{F\'abio R.~Herpich}
\affiliation{Departamento de Astronomia, Instituto de Astronomia, Geof\'isica e Ci\^encias Atmosf\'ericas da USP, Cidade \\ Universit\'aria, 05508-900, S\~ao Paulo, SP, Brazil}

\author{Simone Daflon}
\affiliation{Observat\'orio Nacional - MCTIC (ON), Rua Gal. Jos\'e Cristino 77, São Crist\'ov\~ao, 20921-400, Rio de Janeiro, Brazil}

\author{Carlos E. Barbosa}
\affiliation{Departamento de Astronomia, Instituto de Astronomia, Geof\'isica e Ci\^encias Atmosf\'ericas da USP, Cidade \\ Universit\'aria, 05508-900, S\~ao Paulo, SP, Brazil}
\

\author[0000-0002-0537-4146]{H\'elio D. Perottoni}
\affiliation{Departamento de Astronomia, Instituto de Astronomia, Geof\'isica e Ci\^encias Atmosf\'ericas da USP, Cidade \\ Universit\'aria, 05508-900, S\~ao Paulo, SP, Brazil}

\author{Silvia Rossi}
\affiliation{Departamento de Astronomia, Instituto de Astronomia, Geof\'isica e Ci\^encias Atmosf\'ericas da USP, Cidade \\ Universit\'aria, 05508-900, S\~ao Paulo, SP, Brazil}

\author{Patricia B. Tissera}
\affiliation{Instituto de Astrofísica, Pontificia Universidad Católica de Chile, Av. Vicuña Mackenna 4860, Santiago, Chile}
\affiliation{Centro de Astro-ingeniería, Pontificia Universidad Católica de Chile, Av. Vicuña Mackenna 4860, Santiago, Chile}

\author[0000-0002-4168-239X]{Jinmi Yoon}
\affiliation{Space Telescope Science Institute, 3700 San Martin Dr., Baltimore, MD 21218, USA}
\affiliation{Department of Physics and JINA Center for the Evolution of the Elements, University of Notre Dame, Notre Dame, IN 46556, USA}

\author{Kris Youakim}
\affiliation{Leibniz-Institut f\"ur Astrophysik Potsdam, An der Sternwarte 16, Potsdam 14482, Germany}
\affiliation{Department of Astronomy, Stockholm University, AlbaNova University Center, SE-106 91 Stockholm - Sweden}


\author[0000-0002-4064-7234]{William Schoenell}
\affiliation{GMTO Corporation 465 N. Halstead Street, Suite 250 Pasadena, CA 91107, USA}

\author{Tiago Ribeiro}
\affiliation{Rubin Observatory Project Office, 950 N. Cherry Ave., Tucson, AZ 85719, USA }

\author{Antonio Kanaan}
\affiliation{Departamento de F\'isica, Universidade Federal de Santa Catarina, Florian\'opolis, SC 88040-900, Brazil}


\begin{abstract}

We report photometric estimates of effective temperature, $T_{\rm eff}$, metallicity, [Fe/H], carbonicity, [C/Fe], and absolute carbon abundances, $A{\rm (C)}$, for over 700,000 stars from the Southern Photometric Local Universe Survey (S-PLUS) Data Release 2, covering a substantial fraction of the equatorial Sloan Digital Sky Survey Stripe 82.  We present an analysis for two stellar populations: 1) halo main-sequence turnoff stars and 2) K-dwarf stars of mass $0.58 < M/M_{\odot} <0.75$ in the Solar Neighborhood.  Application of the Stellar Photometric Index Network Explorer (\texttt{SPHINX}) to the mixed-bandwidth (narrow- plus wide-band) filter photometry from S-PLUS produces robust estimates of the metallicities and carbon abundances in stellar atmospheres over a wide range of temperature, $4250 < T_{\rm eff} \textrm{(K)} < 7000$.  The use of multiple narrow-band S-PLUS filters enables \texttt{SPHINX} to achieve substantially lower levels of ``catastrophic failures" (large offsets in metallicity estimates relative to spectroscopic determinations) than previous efforts using a single metallicity-sensitive narrow-band filter.  We constrain the exponential slope of the Milky Way's K-dwarf halo metallicity distribution function (MDF), $\lambda_{10, \textrm{[Fe/H]}} = 0.85 \pm 0.21$, over the metallicity range $-2.5 < \textrm{[Fe/H]} < -1.0$; the MDF of our local-volume K-dwarf sample is well-represented by a gamma distribution with parameters $\alpha=2.8$ and $\beta=4.2$.  
 S-PLUS photometry obtains absolute carbon abundances
 with a precision of $\sim 0.35$\,dex for stars with $T_{\rm eff} < 6500$\,K. We identify 364 candidate carbon-enhanced metal-poor stars, obtain assignments of these stars into the Yoon-Beers morphological groups in the $A$(C)-[Fe/H] space, and derive the CEMP frequencies.

 \end{abstract}


\section{Introduction}

\begin{figure*}
	\centering
	\includegraphics[width=\textwidth, trim=4.0cm 6.0cm 4.0cm 4.0cm, clip]{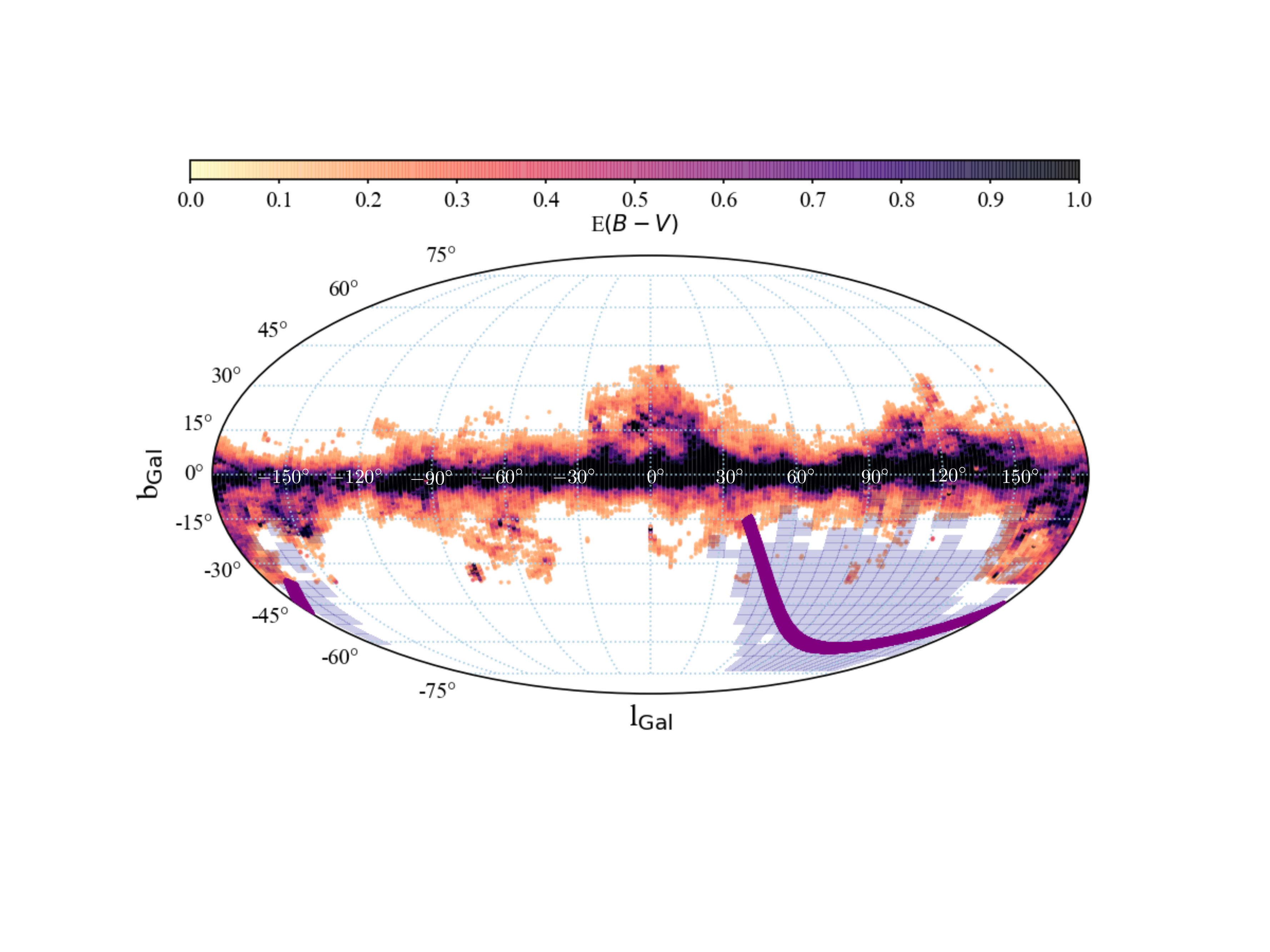}
	\caption{Distribution of the S-PLUS SDSS Stripe 82 sample (\textit{purple}) in Galactic coordinates. The Galactic plane is shown, color-coded by $E(B-V)$, along with the fields sampled by the SEGUE catalog (\textit{blue}) of medium-resolution spectroscopic observations in the Southern Hemisphere (see text for details).\label{fig:galactic_footprint}}
\end{figure*}

The metallicity distribution function (MDF) of the Milky Way (MW) halo and Solar Neighborhood is among the most critical observational constraints for models of Galactic chemical evolution \citep{Salvadori:2008}. The halo MDF is a manifestation of no less than three fundamental properties of the first stellar generations: the stellar initial mass function, the star-formation history, and the chemical-evolution mechanisms at play in the primordial MW, as well as the contribution of stars to the halo from accreted satellite galaxies. It thus is of considerable importance for understanding the early epochs of Galactic evolution.
Historically, \citet{searle1978} were the first to speculate that the formation of the Galactic Halo was much more complex than the rapid, monolithic collapse scenario proposed by the seminal work of \citet{eggen1962}. \citeauthor{searle1978} proposed that the halo was formed by transient protogalactic fragments that fell continuously into dynamical equilibrium with the Milky Way after the collapse of its main structural components
had been completed \citep[see][for a more complete discussion of the subject]{dietz2020}.


In the $\Lambda$CDM paradigm, galactic halos are aggregate structures assembled from the successive merging of a large number of sub-galactic fragments. The MDF of the Galactic halo could thus be considered as the sum of each constituent fragment's MDF \citep{Prantzos:2007}. These fragments are demonstrably distinct from dwarf spheroidal galaxies (dSphs) \textit{as they appear today}, which fail to reproduce the fractions of low-metallicity stars seen in the halo \citep{Helmi:2006, Prantzos:2007}. The observed discrepancies between the Milky Way's halo MDF and elemental-abundance ratios with those of canonical dSph galaxies such as Fornax, Sculptor, Sextans, and Carina present a challenge to models of Galactic evolution that presume a hierarchical assembly of early merging dwarf galaxies \citep{Tolstoy:2009}. Attempts to rectify these differences suppose a pre-enrichment mechanism in the dSph birth environment \citep{Helmi:2006, Salvadori:2008}, although this process is not yet clearly understood, and must necessarily have been quite rapid to have taken place before the onset of star formation. 

In contrast, ultra-faint dwarf galaxies (UFDs) have been demonstrated to possess populations of $\textrm{[Fe/H]} < -3.0$\footnote{Relative abundance ratios used in this work correspond to $\textrm{[A/B]} = \log{(N_A/N_B) - \log{(N_A/N_B)}_{\odot}}$, where $N$ indicates the number density of the given species.} stars \citep{Kirby:2008, Brown:2014}, suggesting that (surviving) UFDs may be more similar to the accreted building blocks that formed the metal-poor tail of the Galactic stellar halo \citep{Frebel:2010, Simon:2019}. However, both the relatively small number of known UFDs, as well as their sparse and faint member stars, present challenges to carry out comprehensive population studies. Observations of the numerous, relatively bright, field stars of the MW's constituent halo populations may thus provide a superior means for probing the Galaxy's progenitor systems.


Chemical-evolution modeling has, in the past, considered the MDF of G-dwarf stars as a primary observational constraint \citep[e.g.,][and references therein]{Caimmi:2008}. These low-mass ($0.8 < M/M_{\odot} < 1.1$) stars possess main-sequence lifetimes older than the disk age, regardless of their metallicity \citep{Bovy:2012}, although some G dwarfs may be massive enough to have already evolved away from the main sequence \citep{Kotoneva:2002}. However, \citet{Fenner:2003} suggests that K-dwarf stars ($0.5 < M/M_{\odot} < 0.8$) are perhaps more directly comparable, as their lower masses render the population less susceptible to stellar evolutionary effects. Models constrained by the G-dwarf MDF suggest a ``two-infall" scenario, with a long timescale for thin-disk formation \citep{Chiappini:2001}. These comparatively long-lived stars present an opportunity to study the chemical distributions of the Solar Neighborhood, while mitigating evolutionary effects and age biases in their selection.

Previous work (e.g., \citealt{Schorck:2009}; \citealt{Li:2010}) relied on comparisons to the halo MDFs obtained by spectroscopic surveys such as the Hamburg-ESO survey (HES; \citealt{Christlieb:2003}). More recently, studies (e.g., \citealt{Carollo:2007}; \citealt{Carollo:2010}; \citealt{Beers:2012}; \citealt{Lee:2017}; \citealt{Yoon:2018:AEGIS}; \citealt{Carollo:2019}; \citealt{Kim:2019}; \citealt{Lee:2019}) have transitioned to spectroscopic surveys of much larger numbers of halo stars, most notably the Sloan Digital Sky Survey (SDSS; \citealt{York:2000}) and its extensions, the Sloan Extension for Galactic Understanding and Exploration (SEGUE; \citealt{Yanny:2009}), the Baryon Oscillation Spectroscopic Survey (BOSS; \citealt{Dawson:2013}), and the Apache POint Galactic Evolution Experiment (APOGEE; \citealt{Majewski:2017}). Nevertheless, such spectroscopic surveys suffer from incompleteness and bias to various degrees, and can be subject to difficult-to-model selection functions.

\begin{figure*}
	\centering

	\includegraphics[width=\textwidth, trim=1.75cm 1.50cm 1.75cm 1.0cm, clip]{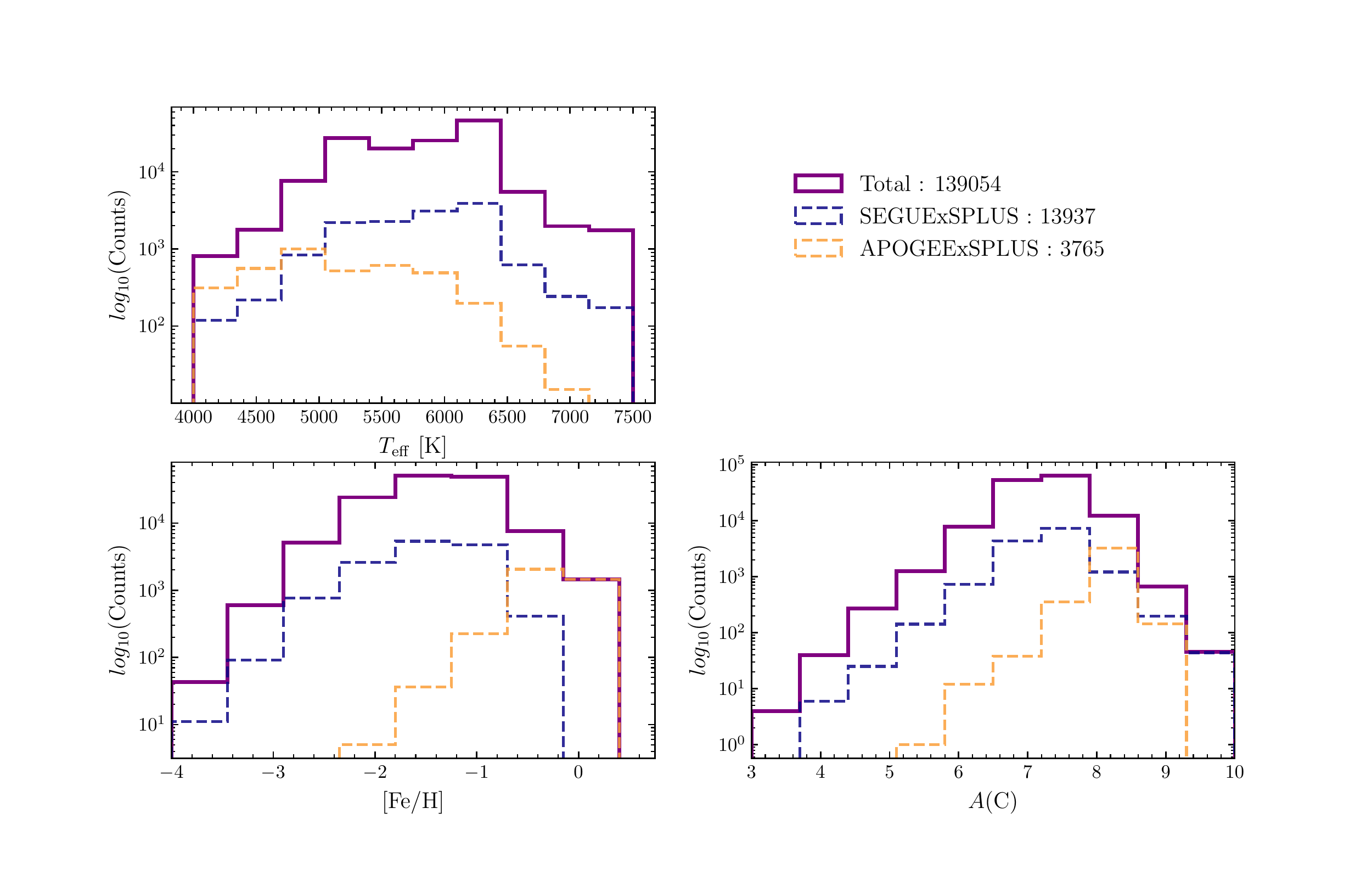}
	
	\caption{Distributions of $T_{\rm eff}$ (top left panel), [Fe/H] (bottom left panel), and $A$(C) (bottom right panel) of the spectroscopic training catalog. The total distribution (\textit{purple}) is shown for each parameter, along with the contributions from the constituent spectroscopic catalogs, SEGUE (\textit{blue}) (which includes legacy SDSS, SEGUE, SEGUE-2, and BOSS) and APOGEE (\textit{orange}), crossed-matched with the S-PLUS (\textit{dashed}) photometric catalog.
	\label{fig:catalog_distro}}
\end{figure*}

While some studies, such as \citet{Bovy:2012} and others (\citealt{Mints:2019}, and references therein), have sought to unravel the effects of spectroscopic-survey selection effects, photometric surveys can substantially reduce these effects, while also sampling a significantly larger fraction of the sky. Using a technique based on broad-band SDSS \textit{ugriz} photometry, \citet{An:2013} found that, when the resulting halo photometric MDF is de-convolved into two Gaussian components (with peaks at [Fe/H] $=-1.7$ and $-2.3$), the more metal-poor component accounts for $20-35$\,\% of the halo population within the distance range $d \sim 5-8$\,kpc from the Sun. It should be noted that this study was limited to main-sequence dwarfs over a restricted range of stellar mass in order to minimize metallicity-related luminosity effects, whereby metal abundance can significantly affect the luminosity of stars and impose a bias on magnitude-limited samples. Based on the ``depth optimized" co-added sample of SDSS Stripe 82 data from \citet{Jiang:2014}, \citet{An:2015} extended the photometric metallicity estimates of stars out to $d \sim 10.5$\,kpc, largely due to inclusion of a fainter $u$-band limit. When their new photometric MDF was de-convolved into multiple populations, with peak metallicities of [Fe/H] $=-1.4$ and [Fe/H] $=-1.9$, the more metal-poor component was found to constitute $\sim35-55\%$ of halo stars in the Solar Neighborhood.

The advent of recent photometric surveys employing mixed-bandwidth photometric filters targeting key absorption features, such as the \ion{Ca}{2} lines\footnote{This technique was first explored in a series of papers beginning with \citet{Anthony-Twarog:1991}, with improved calibrations at lower metallicity by \citet{Anthony-Twarog:2000}.}, have significantly expanding the metallicity sensitivity of photometric surveys. Examples include SkyMapper \citep{Keller:2007} and \textit{Pristine} \citep{Starkenburg:2017}. \citet{Youakim:2020} recently investigated the MDF of main-sequence turnoff (MSTO) stars, extending the metal-poor tail to [Fe/H] $ \sim -3.4$, and constraining the slope of the low-metallicity tail of the halo MDF to $\Delta(\log{N})/\Delta(\textrm{[Fe/H]}) = 1.0 \pm 0.1$. Interestingly, this result is consistent with the very early work from \citet{Beers:1987}, based on only $\sim 130$ metal-poor stars from the HK survey \citep{Beers:1985}.

The Javalambre and Southern Photometric Local Universe Surveys (J-PLUS; \citealt{Cenarro:2019}, S-PLUS; \citealt{Mendes:2019}), are poised to further expand the metallicity sensitivity of photometric studies, as well as to open up the possibility of measuring other elements of importance, such as C and Mg, with the implementation of the Javalambre filter system of 7 narrow-band
and 5 wide-band filters \citep{MarinFranch12}, and a planned combined footprint of over $17,000$\,deg$^2$\footnote{A total of $8,500$\,deg$^2$ and $9300$\,deg$^2$ for J-PLUS and S-PLUS, respectively. J-PLUS and S-PLUS are performed by two 80-cm telescopes called T80/JAST and T80-South, respectively. T80-South and its large-format camera, including the filters (see \href{www.splus.iag.usp.br}{www.splus.iag.usp.br}), located at Cerro Tololo Interamerican Observatory, are a duplicate of the T80/JAST system, installed at Cerro Javalambre (see \href{www.jplus.org}{www.jplus.org} for details).} In \citet{Whitten:2019a}, the Stellar Photometric Index Network Explorer (\texttt{SPHINX}) was demonstrated to be an effective means of utilizing the 12-band filter system used in J-PLUS and S-PLUS to produce reliable estimates of effective temperature and metallicity for stars, based on the relatively small amount of data then available. 

In this paper, we employ \texttt{SPHINX} to study the metallicity ([Fe/H]), carbonicity ([C/Fe]), and absolute carbon abundance ($A$(C)) distributions for two S-PLUS photometric samples of stars from SDSS Stripe 82 -- MSTO stars and cooler main-sequence K-type dwarfs. In Section~\ref{section:Catalogs}, we describe the photometric and spectroscopic catalogs used in this work. In Section~\ref{section:SPHINX}, we detail the structure and training procedure of \texttt{SPHINX} to produce photometric estimates of effective temperature, $T_{\rm eff}$, metallicity, [Fe/H], and absolute carbon abundance, $A$(C)\footnote{$A$(C) $= \log{\epsilon (C)} = \log(N_C/N_H) + 12$, where $N$ represents the number densities of each species.}. We describe the selections made from the S-PLUS Stripe 82 data release to produce a MSTO and a mass-selected local K-dwarf sample in Section~\ref{section:sample_selections}. Finally, we discuss the results of the metallicity and carbonicity distributions and their analyses for both samples in Section~\ref{section:results}, compare with a number of previous efforts, and call particular attention to the issue of ``catastrophic failures" in photometric estimates of metallicity (which our \texttt{SPHINX} analysis of S-PLUS data greatly decreases).  Concluding remarks are provided in Section~\ref{section:conclusions}.


\section{Catalogs}\label{section:Catalogs}

In this section, we describe the S-PLUS Stripe 82 data release employed in this work, and the spectroscopic catalogs used to train and test our derived estimates of stellar $T_{\rm eff}$, [Fe/H], and $A$(C).  Note that the carbon-to-iron ratio, or carbonicity, [C/Fe], is directly obtained from $A$(C) and [Fe/H]:  
[C/Fe] = $A(\rm{C}) - \textrm{[Fe/H]} - A(\rm C)_{\odot}$ (where $A(\rm C)_{\odot}$  = 8.43;  \citealt{Asplund:2009}).

\subsection{S-PLUS Stripe 82}

S-PLUS is an ongoing survey, mapping 9,300\,deg$^2$  of the Southern sky with the Javalambre filter system comprising an optical 12-band narrow- and broad-band photometric system, including five broad-band \textit{ugriz} filters analogous to the SDSS. S-PLUS employs a dedicated 0.8\,m telescope (T80S) located on Cerro Tololo, Chile. The imager, T80Cam, hosts a 9232 $\times$ 9216 10\,$\mu$m-pixel array, with a $1.4^{\circ} \times 1.4^{\circ}$  field-of-view. The public data release\footnote{Available at \href{https://datalab.noao.edu/splus/}{https://datalab.noao.edu/splus/}} of S-PLUS Stripe 82 (\citealt{Mendes:2019}) comprises 170 co-added fields covering a total of 336\,deg$^2$ from $4^h < $ R.A. $ < 20^h$ and $-1.26^{\circ} 
< $ Decl. $ < +1.26^{\circ}$, collected between 2016 and 2018. The S-PLUS raw data are reduced using an adapted early version (number 0.9.9) of the data-processing pipeline \textsc{jype} \citep[see][for details]{2014SPIE.9152E..0OC}, developed by CEFCA's Unit for Processing and Data Archiving, UPAD. This code was designed to reduce data for two surveys, J-PLUS \citep{Cenarro:2019} and J-PAS (Bonoli et al. 2020, submitted), and is based on the photometric pipeline originally developed for the ALHAMBRA survey \citep[see][]{cristobal09, Molino14}. 

 The final magnitude-limited catalog includes $700,000+$ sources down to $r = 21$\ on the AB system. This preliminary catalog is then cleaned of galaxies and other problematic sources (i.e., close double stars, image defects, etc.) via a stellar-probability procedure based on a random forest algorithm \citep{Costa-Duarte:2019}. Only objects with a 90\,\% confidence of being stellar sources (as per DR1) were accepted. The extent of the S-PLUS Stripe 82 sample is shown in Figure~\ref{fig:galactic_footprint}, along with the Galactic plane and the spectroscopic training and validations samples discussed below.

The Javalambre filter system possesses narrow- and intermediate-band filters of particular interest to stellar-parameter determinations. We refer the interested reader to \citet{MarinFranch12}, \citet{Cenarro:2019}, and \citet{Mendes:2019} for an in-depth discussion of this 12-filter system, but highlight three such filters here. The $J0410$ filter ($\Delta \lambda=201$\,\AA) is centered on the H$\delta$ Balmer line ($\lambda = 4102$\AA), which exhibits effective-temperature dependence. The $J0395$ filter ($\Delta \lambda = 103$\,\AA) is centered directly on the \ion{Ca}{2} H \& K lines, which have been used extensively for determinations of metallicity in low- and medium-resolution spectroscopy. Finally, the $J0430$ filter ($\Delta \lambda = 200$\,\AA) is positioned on the CH $G$-band, a useful molecular feature for carbon-abundance determination, also used by previous medium- and high-resolution spectroscopic surveys. 
We note that the Javalambre $J0395$ filter is similar to the Pristine \ion{Ca}{2} H \& K filter \citep{Starkenburg:2017}, which has been successfully used to search for low-metallicity stars. These better isolate the metallicity-sensitive features than the intermediate-band filters from, e.g., SkyMapper \citep[$v$ filter,][]{bessell2011,wolf2018}


\subsection{Spectroscopic Catalogs}

Here we describe the spectroscopic sources used to obtain stellar parameters for training and validation purposes. Figure~\ref{fig:catalog_distro} shows the distribution of stellar atmospheric parameters $T_{\rm eff}$ and [Fe/H], as well as the absolute carbon abundance, $A$(C), available for network training for each of the catalogs described below.

\subsubsection{SEGUE}

The primary catalog employed for network training of the S-PLUS photometric data is derived from the SEGUE \citep{Yanny:2009} and SEGUE-2 (Rockosi et al., in prep.) surveys.  We also include observations taken as part of the SDSS Legacy Survey \citep{Abazajian:2009}, as well as from BOSS \citep{Dawson:2013}. For simplicity of notation, we refer to the complete set of these data as the SEGUE sample, since it is the dominant contributor. By application of the SEGUE Stellar Parameter Pipeline (SSPP, \citealt{Lee:2008a, Lee:2008b}), atmospheric parameters and estimates of [Fe/H] and $A$(C) are obtained for the $\sim 300,000$ stars in common with available spectroscopy. The S-PLUS Stripe 82 sample was then cross-matched with this catalog using a search radius of 5\arcsec, resulting in 23,882 stars with available S-PLUS photometry and SEGUE/SSPP stellar atmospheric parameters. We utilize this spectroscopic sample for comparison to our photometric measurements with S-PLUS. 

\subsubsection{APOGEE}

To extend the metallicity range of our photometric training sets, in particular for more metal-rich stars, we include a subset of stars from APOGEE \citep{Majewski:2017}. APOGEE is a high-resolution ($R \sim 22,500$), near-infrared ($1.5-1.7$\,$\mu m$) stellar spectroscopic survey performed at both the Apache Point Observatory in New Mexico, USA, using the APOGEE-North spectrograph on the 2.5-meter Sloan Foundation Telescope, and, as of Dec 2019, at Las Campanas Observatory, using the essentially identical APOGEE-South spectrograph. Chemical abundances produced with the APOGEE Stellar Parameters and Abundances Pipeline (ASPCAP; \citealt{ASPCAP}) serve as the primary catalog for our sources with $\textrm{[Fe/H]} > -0.5$. While the SEGUE  and APOGEE cross-matches have a similar number of stars with $\textrm{[Fe/H]}>-0.5$ ($\sim5,000$), the higher resolution and typically much higher signal-to-noise ratios of the APOGEE spectra are expected to yield superior results to those of SEGUE for $\textrm{[Fe/H]}>-0.5$ \citep{Meszaros:2013}. We note a small systematic offset, $\textrm{[Fe/H]}_{\rm SEGUE} - \textrm{[Fe/H]}_{\rm APOGEE} = -0.09$\,dex, between the metallicity reported by ASPCAP and the biweight metallicity estimate adopted from the SSPP. We adjust the ASPCAP metallicities to remove this offset in the master catalogs employed in the [Fe/H] and $A$(C) training exercise. A larger offset is seen in the A(C) values, $\textrm{A}(C)_{\rm SEGUE} - \textrm{A}(C)_{\rm APOGEE} = -0.21$\,dex. This offset is also corrected for, and we note that stars from APOGEE represent only $\sim$15\% of total stars in the training samples.

\begin{figure*}
	\centering
	\includegraphics[width=\textwidth, scale=0.82, trim=0.0cm 0.55cm 0.0cm 0.00cm, clip]{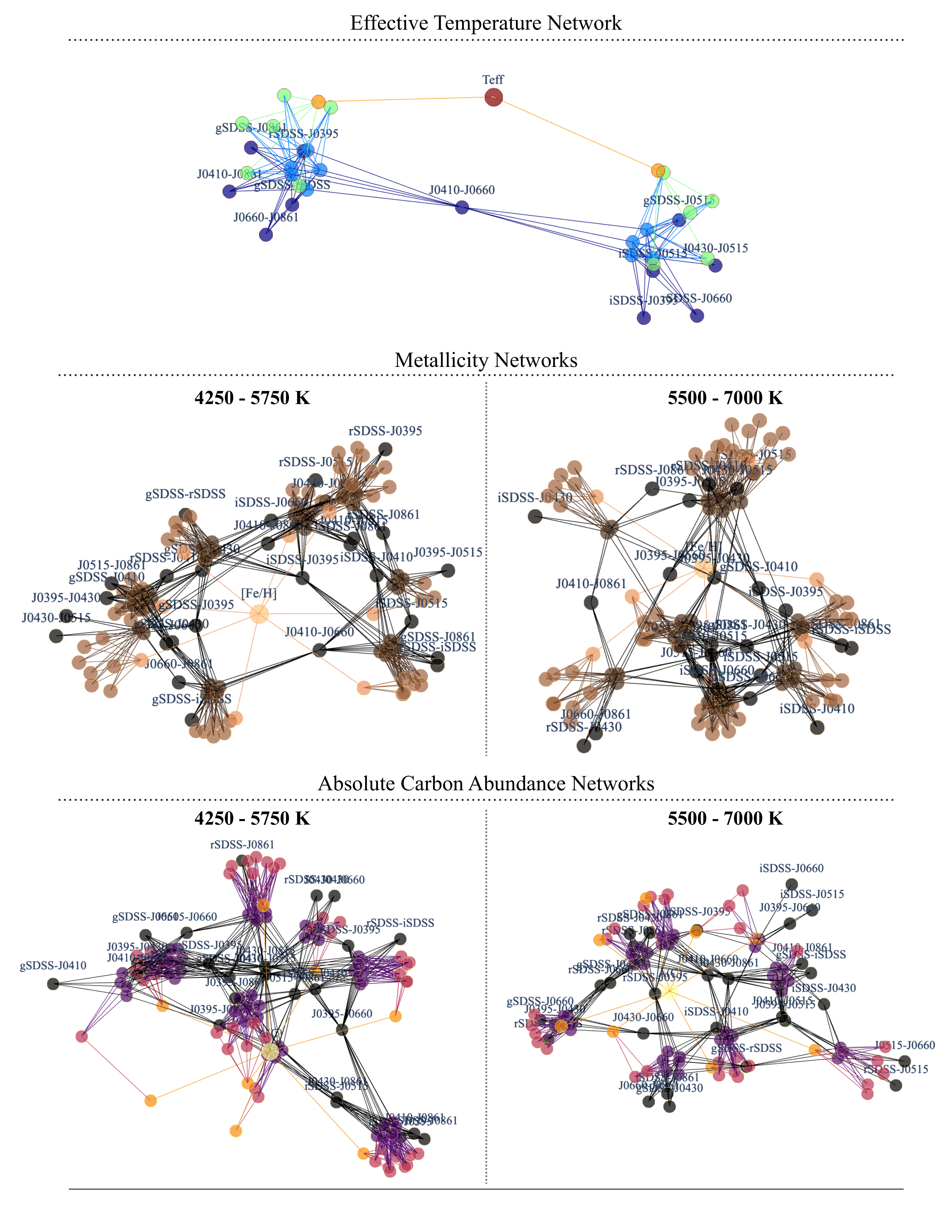}
	
	\caption{Architecture of the generative training process for the effective temperature ($T_{\rm eff}$) network (upper panel), the metallicity ([Fe/H]) networks (split into two $T_{\rm eff}$ regimes, as indicated in the legend), and the absolute carbon ($A$(C)) network, split into the same $T_{\rm eff}$ regimes. Black dots indicate the photometric color inputs, and remaining colors indicate successive layers in each model. See text for more details.
	For a description of the filters, see 
	\href{http://www.splus.iag.usp.br/instrumentation/ } {http://www.splus.iag.usp.br/instrumentation/ } \label{fig:NETWORK_DIAGRAMS}}
\end{figure*}

\subsubsection{Sample Contamination}

We address possible contamination of our photometric samples from S-PLUS by background quasars via  cross-matching with the Large Quasar Astrometric Catalog \citep{Gattano:2018}. This catalog consists of nearly all known quasars, with the addition of recent identifications in the SDSS DR12Q, as well as $Gaia$ DR1 \citep{Gaia:DR1}. We find $9,238$ matches in the S-PLUS Stripe 82 sample, which were removed from all subsequent analysis.

We additionally note the potential for contamination by cool white dwarfs in our samples. While the vast majority of white dwarfs that could be present in our data are substantially hotter than the temperature selection implemented for both the MSTO and K-dwarf samples, cooler ($T_{\rm eff} < 10000$) white dwarfs can occupy a similar $(g-i)_0$ range as the MSTO sample. White dwarfs possess significantly higher surface gravity than MSTO or K-dwarf stars, $\log{g}>7$ \citep{Eisenstein:2006}. They can thus be effectively excluded by the color restriction $(u-g)_0 > 0.6$, as $(u-g)_0$ is a known surface-gravity dependent feature, due to the Stark pressure broadening (and overlap) of Balmer lines in the near-UV region. Finally, white dwarfs are intrinsically faint, which severely limits the distance range within which white dwarfs can populate our samples. We conclude that white dwarfs are not a significant source of contamination on our sample selections.

The main samples selected in this work consist of warm MSTO stars and cool K-dwarf stars. We estimate the extent to which these samples are contaminated by giants from an absolute magnitude ($M_g$) calibration using the subset of the S-PLUS Stripe 82 sample cross-matched with $Gaia$ DR2 \citep{Gaia}, using a search radius of 5\arcsec. Of the 36,630 stars with acceptable quality S-PLUS photometry and \textit{Gaia} astrometry, $\sim 14\%$ possess a $M_g >4$. Giant contamination is primarily a concern for the K-dwarf sample, where the contamination is determined to be $\sim 25$\,\%. We address the measures taken to mitigate this factor in the sample selection in \S 4 below.

SDSS Stripe 82, was, by design, frequently revisited during the course of the various SDSS spectroscopic surveys. Thus, our spectroscopic catalogs were internally cross-matched with a search radius of 2\arcsec\ to identify duplicate observations. In the event of a duplicate observation, the spectrum with the highest S/N ratio is chosen. Approximately 23\% of SEGUE sources were found to have duplicate spectroscopic observations. These stars were included for training purposes, as they are legitimate sources. For the determinations of chemical-abundance distributions described in \S 5 below, these stars were removed.

\subsection{Photometric Calibration of S-PLUS Observations}

The sample used in this paper is a revision of the S-PLUS DR1, where a new calibration of the data was employed. The technique used for calibration is the same as that of DR1, described in \citet{Mendes:2019}.  It relies on fitting stellar templates to a reference catalog (in this case, SDSS; \citealt{Ivezic:2007}) in order to predict the S-PLUS calibrated magnitudes for a selection of stars. Then, the zero-points are derived from the differences between the instrumental and the predicted magnitudes. The main differences between the S-PLUS DR1 and the revised DR1 sample used here are that (i) in the latter an aperture correction was applied to the 3\arcsec\ magnitudes before calibration, and (ii) the revised calibration used a library of theoretical stellar spectra \citep{Coelho:2014}, instead of the empirical library used for DR1. These changes, as well as a more detailed description of the calibration process, are discussed in the upcoming S-PLUS DR2 paper (Almeida-Fernandes et al., in prep.). Extinction corrections were performed on this catalog, using $E(B-V)$ values from \citet{Schlafly:2011}, based on the \citet{Schlegel:1998} dust map, and extinction coefficients from \citet{LopezSanjuan:2019}. This dust map is shown in comparison with the S-PLUS Stripe 82 footprint in Figure~\ref{fig:galactic_footprint}.\\
\\

\section{Processing with \texttt{SPHINX}}\label{section:SPHINX}

Here we describe the basic operations and training procedures of \texttt{SPHINX}. We refer the interested reader to the more complete descriptions in \citet{Whitten:2019a}.

\subsection{Generative Network Training}

\texttt{SPHINX} is essentially an Artificial Neural Network (ANN) consensus procedure, consisting of a network of ANN sub-units, each with a unique combination of S-PLUS mixed-bandwidth photometric colors. Each ANN sub-unit is generated in a stochastic manner, whereby the number of nodes, the activation function, and the solution algorithm are randomly assigned, subject to a number of conditions. First, the [Fe/H] and $A$(C) ANN sub-units are restricted to two hidden layers; the addition of more hidden layers produced no notable improvements in the model's predictive power. Each ANN sub-unit is designated one of the following activation functions at random: identity, hyperbolic tangent, logistic, and the rectified linear unit. We additionally randomize the solution function implemented in each sub-unit: the optimized gradient-based algorithm \textit{Adam} \citep{ADAM}, Limited-memory Broyden-Fletcher-Goldfarb-Shanno algorithm (L-BFGS), and the conventional Stochastic Gradient Descent (SGD).  The number of color inputs to each ANN sub-unit is fixed at six, with each ANN sub-unit receiving a unique combination of the available colors. Finally, for each respective stellar-parameter network ($T_{\rm eff}$, [Fe/H], and $A(\textrm{C})$), we force the inclusion of that parameter's most sensitive narrow-band filter in all of its constituent ANN sub-units, corresponding to $J0410$, $J0395$, and $J0430$, for the $T_{\rm eff}$, [Fe/H], and $A$(C) networks, respectively.

All of the ANN sub-units are then trained on $\sim2,000$ sub-samples from the SEGUE calibration set, including APOGEE, selected by uniformly sampling the parameter in question. The effective temperature network is trained across the range $4000 < T_{\rm eff} \textrm{(K)} < 7500$, the metallicity network is trained across $-3.5 < \textrm{[Fe/H]} < +0.2$, and the absolute carbon abundance network is trained across $5.00 < A{\rm (C)} < 8.75$. We found during training that $A$(C) estimates were infeasible for stars with $T_{\rm eff} > 6500$\,K, due to the minimal impact of carbon on the filter set for such warm stars. This limitation  prevents consideration of the carbon abundances for our full MSTO sample, but no such difficulty exists over the effective temperature range of our K-dwarf sample.

Validation scores are determined for each network according to the average residual, evaluated with an independent validation set. For each parameter network, the top eight ANN sub-units are selected from the ensemble. These optimized ANN sub-units are shown in Figure~\ref{fig:NETWORK_DIAGRAMS} for [Fe/H] and $A$(C), where we visualize the aggregate network using the Kamada-Kawai algorithm \citep{KAMADA1989}. These validated ANN sub-units form the basis of the consensus method for parameter determination used throughout this work, where final parameter estimates are the average of the ANN sub-unit estimates, weighted by the validation scores. For both metallicity and absolute carbon abundance, separate networks are constructed and trained for cool $4200 < T_{\rm eff} {\rm(K)} < 5750$ and warm $5500 < T_{\rm eff} {\rm(K)} < 7000$ stars. Stars are directed to each effective temperature partition using the \texttt{SPHINX} $T_{\rm eff}$ estimate. For stars in the intermediate region, $5500 < T_{\rm eff}{\rm(K)} < 5750$, estimates from both networks are merged, and the average of the two estimates is taken.

The uncertainties reported for the [Fe/H] and $A{\rm(C)}$ estimates are based on the scatter amongst the individual ANN sub-units in each network. While representative of the uncertainty of the estimation procedure, these values were found to be quite low, and are unlikely to reflect the underlying scatter seen in the spectroscopic validation stars, $\sigma(\textrm{[Fe/H]}) = 0.25$\,dex and $\sigma(A{\rm(C)})=0.35$\,dex, respectively. We obtain a final estimate of the uncertainty in these estimates by adding the standard deviation of the spectroscopic residuals and the uncertainties reported from \texttt{SPHINX} in quadrature. We discuss validations of the $A$(C) determinations from \texttt{SPHINX} in the following subsection.

\subsection{Absolute Carbon Validation}\label{section:ac_validation}

\begin{figure*}
	\centering

	\includegraphics[trim=2.0cm 1.50cm 3.0cm 0.0cm, clip, width=\textwidth]{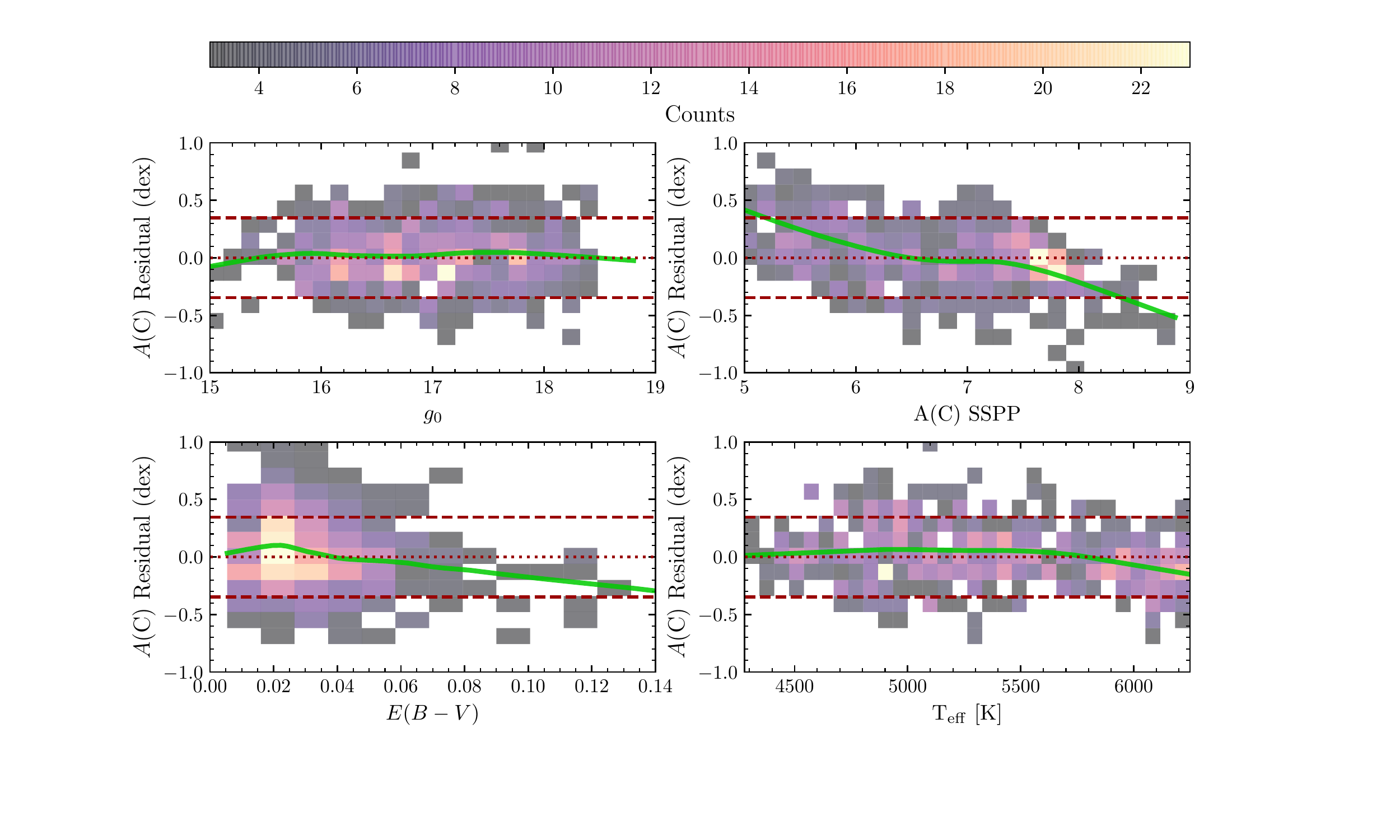}
	\caption{$A$(C) residual trends in validation samples with \texttt{SPHINX}. The difference between the \texttt{SPHINX} photometric $A$(C) estimate and the spectrscopically determined values are shown plotted against the $g$ magnitude (top left panel), the spectroscopic $A$(C) value (top right panel), the reddening, $E(B-V)$ (bottom left panel), and the photometric effective-temperature estimate (bottom right panel). Trend-lines (green panel) are determined via locally weighted regression. \label{fig:ac_validation}}
\end{figure*}

The effective temperature and metallicity determinations from \texttt{SPHINX} were validated in \citet{Whitten:2019a}, and we refer the interested reader to the details therein. Here, we investigate the $A$(C) determinations developed in this work. In Figure~\ref{fig:ac_validation}, the $A$(C) residuals -- defined as the difference between the photometric value determined with \texttt{SPHINX} and the adopted spectroscopic estimate -- are shown plotted against four parameters of particular interest. The $A$(C) residual is shown as a function of the $g$-band magnitude in the top left panel of Figure~\ref{fig:ac_validation}, along with the spectroscopic $A$(C) estimate (top right panel), the reddening estimate $E(B-V)$ (bottom left panel), and finally the photometric effective temperature produced with \texttt{SPHINX} (bottom right panel). For the $g$-band magnitude, reddening, and effective temperature, any trends seen in the $A$(C) estimates are well within the standard deviation in the $A$(C) residuals, $\sigma(A({\rm C})) = 0.35$\,dex.

We note the apparent trend in the $A(C)$ residuals with the reddening, $E(B-V)$. However, the majority of stars in our samples possess reddening values less than $E(B-V) < 0.1$, with a maximum of $E(B-V) <0.15$. As this trend is within the standard deviation of the $A$(C) residuals, we do not impose a correction on the photometric $A$(C) values determined with \texttt{SPHINX}. This is partly due to the nature of the $E(B-V)$ values utilized in this work, namely the \citet{Schlafly:2011} recalibration of \citet{Schlegel:1998}, which do not take distance into account. While 3-D estimates have been produced to take into account the line-of-sight distance, for instance the work of \citet{Green:2019} based on {\em Gaia}, Pan-STARRS, and 2MASS, these estimates are currently limited to declinations with $\delta >-30^{\circ}$, and ultimately rely on high-quality distance estimates, which can introduce a significant amount of uncertainty.
Finally, a trend in the $A$(C) residual with the spectroscopically determined $A$(C) value is seen in Figure~\ref{fig:ac_validation}. Below $A({\rm C})<6.0$, photometric estimates are seen to over-estimate the spectroscopic value. Above $A({\rm C}) > 8.0$, the photometric values under-estimate the carbon abundance, compared with the spectroscopically determined values. Further investigation revealed no clear origin for this trend, beyond the difficulties inherent in carbon estimation, namely the nascent saturation of carbon features beyond $A({\rm C}) > 8.5$, as discussed in \citet{Yoon:2020}. We emphasize the preliminary nature of the photometric carbon abundances produced in this work, where we implement a threshold for absolute carbon estimates of $A({\rm C}) < 8.5$, and reserve further refinements of the photometric $A$(C) determinations to future studies.

\section{Sample Selection}\label{section:sample_selections}

Here we describe the stellar isochrones we employ, and the various criteria adopted to select MSTO stars and K-dwarf stars from the S-PLUS Stripe 82 data release.\\
\\
\subsection{Stellar Isochrones}

The stellar isochrones used in this work were produced from the Yonsei-Yale (YY; \citealt{YY:2004}) models, which take into account helium diffusion and convective-core overshoot. Isochrones were generated in the range [Fe/H] $= [-4.0, 0.0]$ and ages of 10, 11, and 12\,Gyr, where we linearly interpolated between models as necessary. For all models, the $\alpha$-element abundance ratio was set to $[\alpha/\textrm{Fe}] = +0.4$. These models were transformed to equivalent SDSS photometry via the Johnson-Cousins to SDSS calibrations of \citet{Jordi_2006}. 

For validation purposes, we make use of distances computed with \texttt{StarHorse} \citep{starhorse}, which combines parallax estimates and optical photometry from \textit{Gaia} DR2 with photometric catalogs from Pan-STARRS, 2MASS, and AllWISE, to produce Bayesian distance estimates for 265 million stars brighter than $G=18$\footnote{\href{https://data.aip.de/projects/starhorse2019.html}{https://data.aip.de/projects/starhorse2019.html}}. Distance estimates from \texttt{StarHorse} were limited to those for which the fractional posterior uncertainty was $(d_{84} - d_{16})/d_{50} < 0.3$, where $d_{84}$ and $d_{16}$ are the 84th and 16th percentiles, respectively.

\begin{figure}
	\centering
	\includegraphics[width=\columnwidth, trim=0.5cm 0.00cm 0.0cm 0.50cm, clip]{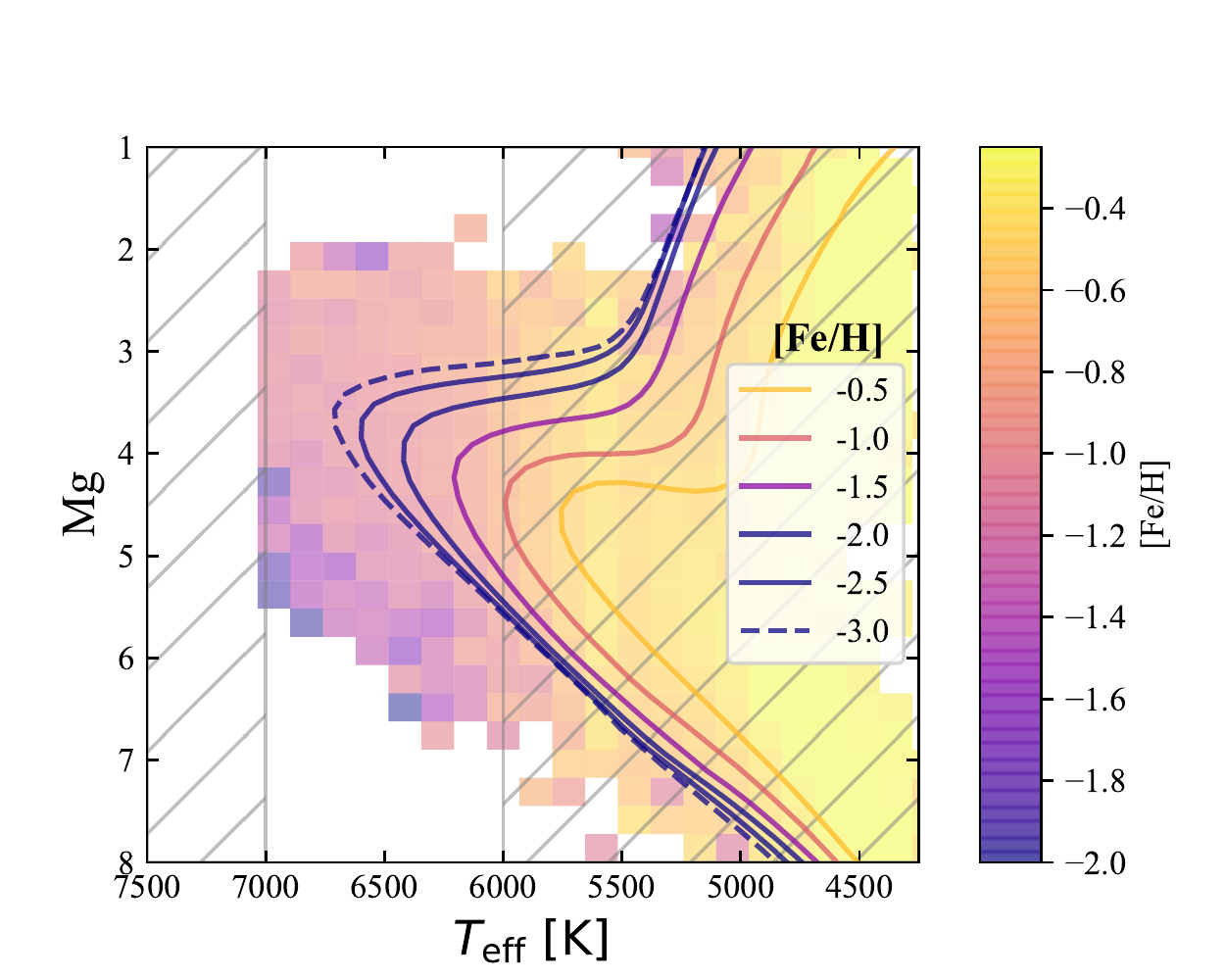}
	
	\caption{Comparison with the Yonsei-Yale isochrones used for the selection of MSTO stars. The \texttt{SPHINX} metallicity estimates are represented by color, where the selection region for MSTO stars, $6000 < T_{\rm{eff}}\textrm{(K)} < 7000$, is denoted by the vertical gray lines. \label{fig:MSTO_selection}}
\end{figure}

\begin{figure*}
	\centering
	\includegraphics[width=\textwidth, trim=0.0cm 0.00cm 0.50cm 0.0cm, clip]{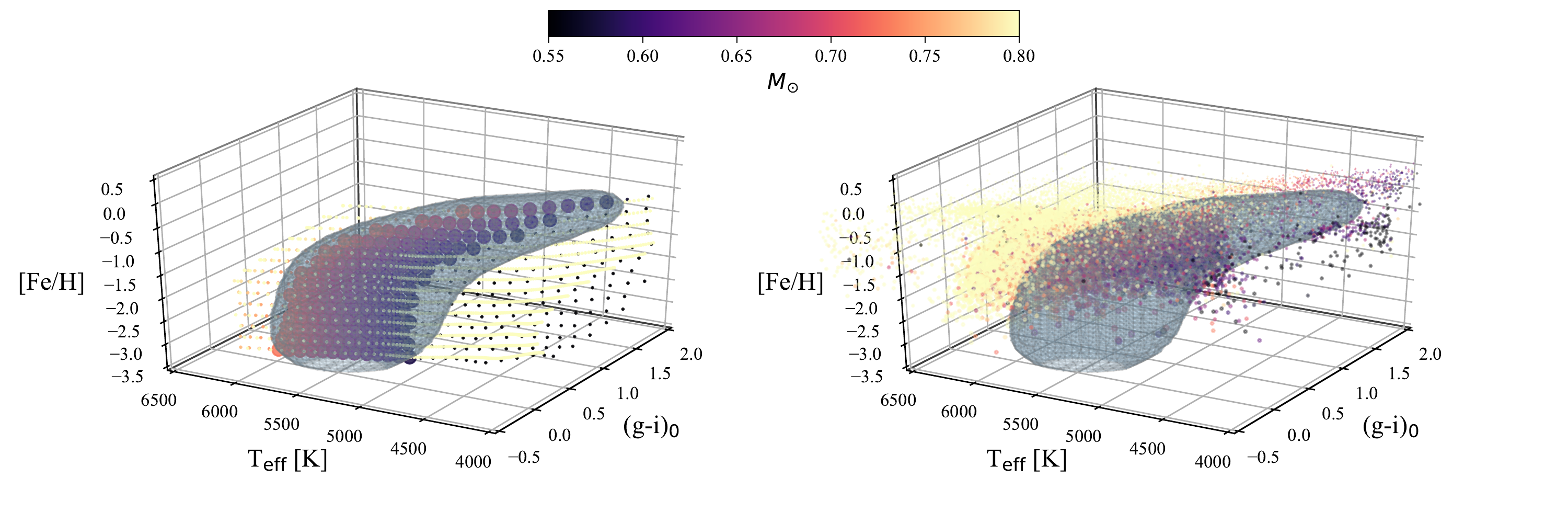}
	\caption{Mass-based selection criteria for the S-PLUS Stripe 82 sample. The left panel shows the stellar isochrones from \citet{YY:2004}, corresponding to the mass range $0.58 < M/M_{\odot} < 0.75$, color-coded by mass. The resulting SVM decision boundary is shown in the right panel, along with the S-PLUS Stripe 82 sample with \texttt{StarHorse}-based mass estimates\label{fig:SVM_MASS_SELECTION}. The gray region in both plots represents the decision boundary, within which stars are selected. See text for details.}
\end{figure*}

\subsection{Main-Sequence Turn-off Selection}

We select MSTO stars from the S-PLUS Stripe 82 catalog primarily on the basis of effective temperature, with an additional $(g-i)_0$ selection. Figure ~\ref{fig:MSTO_selection} shows the temperature-selection window, corresponding to 12\,Gyr isochones. 

For the calibration of our selection function, absolute magnitudes, $M_g$, are determined using distance estimates obtained by \texttt{StarHorse}. We select MSTO stars within the range $6000 < T_{\rm eff} \textrm{(K)} < 7000$. This sample is additionally subjected to a restriction on the vertical distance from the Galactic plane, $|Z| > 3$\,kpc. As distance estimates from \texttt{StarHorse} are not available for the majority of the stars in our sample, the spatial coordinates are determined via an approximate distance modulus, whereby an absolute magnitude $M_{g} = 4$ for the MSTO sample was uniformly adopted. Additionally, a Galactic latitude limit of $|b| > 30^{\circ}$ was applied to mitigate contamination from the disk system. Of the 88,469 stars with acceptable S-PLUS photometry and \texttt{SPHINX} metallicities, our final MSTO sample consisted of 6,360 stars. An identical selection was performed on our SEGUE spectroscopic catalog to produce a sample of 4,038 stars.

\subsection{Mass-Based K-Dwarf Selection}\label{sec:dwarf_selection}

In addition to the effective-temperature selection for MSTO stars, we conduct a selection on mass for main-sequence K-dwarf stars in the range $0.58 < M/M_{\odot} < 0.75$. This is done for several reasons. 
First, this region of stellar mass corresponds to a color-space selection in $(g-i)_0$ and $(u - g)_0$ that is effectively independent of age. Stars selected on mass in this manner are therefore unbiased towards any particular age.
Secondly, we can compare the validity of our chemical-abundance distributions to the work of \citet{An:2015}, which implemented a similar mass range, $0.65 < M/M_{\odot} < 0.8$.  Lastly, this mass range corresponds to an 
effective-temperature range ($4250 < T_{\rm eff}\textrm{(K)} < 6250$) that permits confident photometric estimations of stellar carbon abundance, in addition to metallicity.

Our mass-based selection comprises a three-dimensional criteria: $T_{\rm eff}$, $(g-i)_0$\footnote{The reddening treatment is the same as the one described in Section~\ref{section:ac_validation}.}, and [Fe/H]. We develop the selection region on the basis of 12\, Gyrstellar isochrones, in conjunction with a Support Vector Machine (SVM), in order to robustly determine the 3-D contour region populated by the stars in this mass range. 

To simulate the expected uncertainty inherent in the spectroscopic SEGUE and photometric catalogs, we inject the YY stellar isochrones with random noise according to the following: $\sigma(T_{\rm eff}) = 100$\,K, $\sigma(g - i)_0= 0.2$\,mag, and $\sigma(\textrm{[Fe/H]}) = 0.3$\,dex. These noise-injected isochrones form the basis of our decision function. 

The isochrones are then reduced to a mass range corresponding to $0.57 < M/M_{\odot} < 0.75$. To train our SVM on the decision boundary, we first subject the isochrone parameters $T_{\rm eff}$, $(g-i)_0$, and [Fe/H] to Principal Component Analysis (PCA), in order to maximize the statistical variance along each of the transformed principal axes, while centering the distribution and scaling the inputs to unity variances.

We adopt a similar SVM training methodology to that described in \citet{Whitten:2019b}, namely a radial basis function. The width of the radial basis function, $\gamma$, is set to the variance of the training set. We additionally weight the cost function by the inverse of the class frequencies in our training set, consisting of $\sim 7,900$ positive examples, and $\sim 7,600$ negative samples. Negative examples were assembled by embedding the positive mass-selected isochrone sample in a surrounding mesh. Surrounding the positive examples from the isochrone ensures a closed contour following SVM training.

The resulting decision boundary for the mass selection is shown in Figure~\ref{fig:SVM_MASS_SELECTION}. In the left panel, the YY isochones that formed the basis of the decision boundary are shown in gray.
In the right panel, we overlay the sub-sample of the S-PLUS Stripe 82 catalog with mass estimates from \texttt{StarHorse}.

\begin{figure}
	\centering
	\includegraphics[width=\columnwidth, trim=0.0cm 0.00cm 2.00cm 0.00cm, clip]{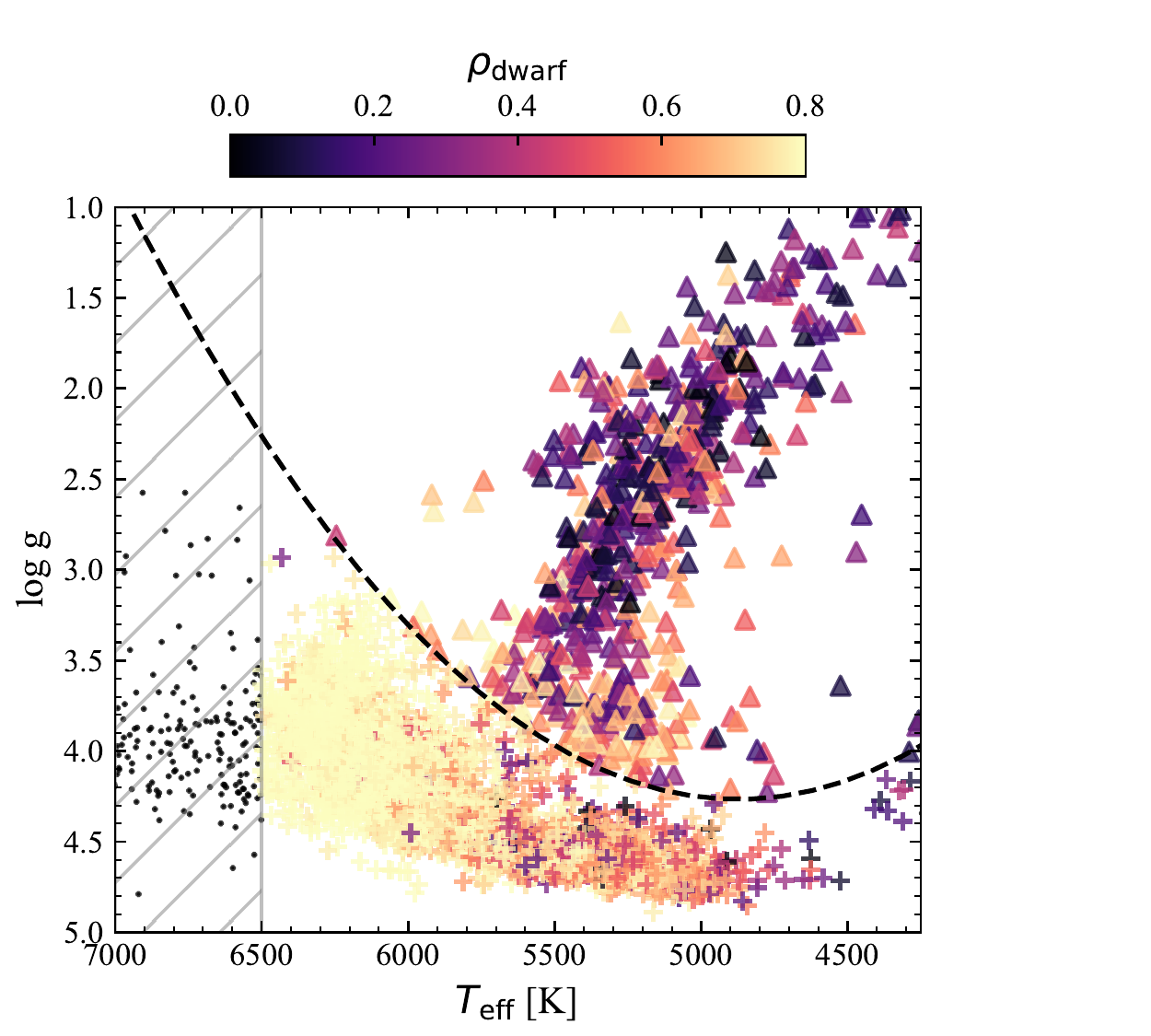}
	
	\caption{Hertsprung-Russell diagram with surface-gravity classifications from the Random Forest Classifier. Dwarf examples are shown as `+' symbols, and giants as triangles, with the resulting dwarf probabilities represented by color. The dashed line corresponds to the discrimination region between dwarfs and giants used for classification. The dashed region indicates stars beyond the temperature range of the surface-gravity classification, provided for reference nonetheless. 
	\label{fig:RFC_logg}}
\end{figure}

We note that, particularly for higher values of metallicity, [Fe/H] $>-1.0$, some contamination from giant-branch stars is present within the selection region. Using the subset of the S-PLUS Stripe 82 with spectroscopic surface-gravity estimates from SEGUE, we estimate the contamination level to be $\sim 25$\%. However, \citet{Guillaume:2019} demonstrated the capability of Random Forest Classifiers (RFCs) to discriminate between dwarfs and giants with a combination of the Canada-France Imaging Survey \citep{Ibata:2017}, Pan-STARRS 1, and \textit{Gaia} photometry. We adopt a similar procedure, based on surface-gravity sensitive photometric S-PLUS colors: $(u\rm{JAVA} - g\rm{SDSS})_0$, $(g\textrm{SDSS}-r\textrm{SDSS})_0$, $(J0378-J0515)_0$, and $(J0660-J0861)_0$. We train this RFC using dwarf ($N=3,139$) and giant ($N=632$) examples, based on the spectroscopic surface-gravity estimates from SEGUE. To address the class imbalance between dwarfs and giants during the training procedure, we implement inverse-frequency class weighting.
The dwarf/giant classification probabilities are shown for an independent validation set in Figure~\ref{fig:RFC_logg}. The dashed line represents the boundary assigned between dwarf and giant classifications for training. 
Only stars with $\rho_d> 0.5$ and $\rho_g <0.5$ are accepted as dwarfs, where $\rho_d$ and $\rho_g$ represent the RFC-assigned likelihood of dwarf and giant classification, respectively.
Following the RFC giant rejection, contamination is reduced from $\sim  25$\% to $\sim 15$\%. 
We emphasize that this RFC contamination removal preserves the completeness of the sample; only 9\% of true dwarf stars are lost following the RFC rejection. Of the 111,815 stars remaining after the SVC mass-selection procedure, 19,735 were rejected by the RFC  dwarf-selection procedure. Given the initial giant-contamination estimate of $\sim 25$\%, the estimated total number of remaining giants is $\sim 28,000$, consistent with a final contamination rate of $\sim 10$\% following RFC rejection.
Our final sample of K-type dwarfs comprises 52,035 stars.

\section{Results and Discussion}\label{section:results}

In this section, we briefly describe the methodology used to study the metallicity and absolute carbon abundance distributions of our MSTO and K-dwarf samples, followed by a presentation and discussion of the results.

\subsection{The MSTO Distribution}

The MSTO sample includes stars that are sufficiently distant ($3 < d~(\textrm{kpc}) < 9$) to comprise stars from the inner- and outer-halo populations, with a likely additional contribution from the thick-disk system. We model the MSTO MDF as the sum of individual Gaussian components, employing Gaussian Mixture Modeling (GMM):
    
\begin{equation}
\rho(z | \mu_i, \sigma_i) = \sum_{i=1}^{K} \phi_{i}\mathcal{N}(z, | \mu_i, \sigma_i).
\end{equation}  

\noindent Here, we use $z$ to indicate the metallicity, [Fe/H]. Individual Gaussian functions, $\mathcal{N}$, are weighted according to the variable $\phi$. We estimate the appropriate weights, central metallicities, $\mu_i$, and scales, $\sigma_i$. 
    
\subsection{Reproducing Distribution Families for the K-Dwarf Sample}

The K-dwarf sample from S-PLUS covers a substantially smaller range of distance than the MSTO sample, $1 < d \textrm{(kpc)} < 6$. Although there are certainly inner- and outer-halo stars represented in the volume we consider, we expect the majority of stars to come from the thick-disk system (including members of both the canonical thick disk and the metal-weak thick disk).

To constrain the shape of the K-dwarf sample MDF, we employ two distinct distribution families: the exponential distribution, and the gamma distribution. Most commonly used to describe the extent of the low-metallicity tail, the exponential probability distribution function can be written as:

\begin{equation}\label{eq:expo_e}
\rho_e(z | \lambda_e) = N(\lambda_e | z_0, z_f)\cdot e^{-\lambda_e |z|}.
\end{equation}

\noindent The exponential slope is equivalent to $\lambda = \dot{\rho_e}$. Here, $z_0$ and $z_f$ represent the metallicity limits used for normalization. We include the normalization factor, $N$, to account for the finite range of the metallicity estimates accessible with our photometric technique:

\begin{equation}
N_{*}(\boldsymbol{\theta} | z_0 , z_f) = \left[ \int_{z_0}^{z_f} \rho(z | \boldsymbol{\theta}) dz \right]^{-1}.
\end{equation}

\noindent Here, $\boldsymbol{\theta}$ represents the parameters associated with the model. The range over which photometric metallicites are estimated in this work is $-3.5 < \textrm{[Fe/H]} < 0.0$, although we additionally consider more narrow ranges.

For ease of comparison with similar previous studies, we additionally provide the following form of the exponential distribution:

\begin{equation}\label{eq:expo_10}
\rho_{10}(z | \lambda_{10}) = N(\lambda_{10} | z_0, z_f)\cdot 10^{-\lambda_{10} |z|}.
\end{equation}

\noindent Note that the exponential slopes are related by a simple change of base: $\lambda_{10} = \lambda_e \log_{10}{e}$.

In addition, we explore the gamma distribution function. We chose this function for its flexibility in accommodating a diverse range of distribution shapes, and for simplicity in the interpretation of its parameters. The gamma distribution function, $\rho_{\Gamma}$, is:

\begin{equation}\label{eq:beta_pdf}
    \rho_{\Gamma}(z | \alpha, \beta) = N_{\Gamma}(\alpha, \beta | z_0, z_f) \frac{\beta^{\alpha}}{\Gamma(\alpha)} \cdot |z|^{\alpha - 1}e^{- \beta z}.
\end{equation}

\noindent Here, the shape parameter, $\alpha$, governs the skewness, and the inverse-scale parameter, $\beta$, describes the extent of the distribution. We apply an absolute value on the metallicities to transform to the applicable domain $|z| \in (0,\infty)$. 

While both forms of the exponential distribution involve the determination of a single parameter, $\lambda$, the gamma distribution involves constraining both $\alpha$ and $\beta$. Thus, it becomes important to consider the covariance of both parameters. We implement a Bayesian Maximum Likelihood Estimation (MLE) approach in order to estimate the optimal combination of ($\alpha, \beta$). Assuming independently sampled data, the model likelihood is simply the product of the individual probabilities:

\begin{equation}
\mathcal{L}(\alpha, \beta) = \prod_{i}^{N} \rho_{\Gamma}(z_{i} | \alpha, \beta) \rho(\alpha, \beta).
\end{equation}

\noindent 
All priors are taken to be uniform, and the logarithm of the likelihood is sampled across the posterior space using the \texttt{emcee} \citep{emcee} implementation of the Markov Chain Monte Carlo (MCMC; \citealt{Goodman:2010}) method. 
As the range of our photometric metallicity sensitivity is limited, we ensure that all PDFs are renormalized according to the range [Fe/H] $=[-3.5, -0.5]$. The renormalization is itself a function of the model parameters $\alpha$ and $\beta$, thus we implement a linear grid interpolation for reference during the MCMC determinations. This grid interpolation samples the values of $\alpha$ and $\beta$ in increments of $0.01$.

\begin{figure*}
	\centering
	\includegraphics[width=\textwidth, trim=1.50cm 0.00cm 1.50cm 0.50cm, clip]{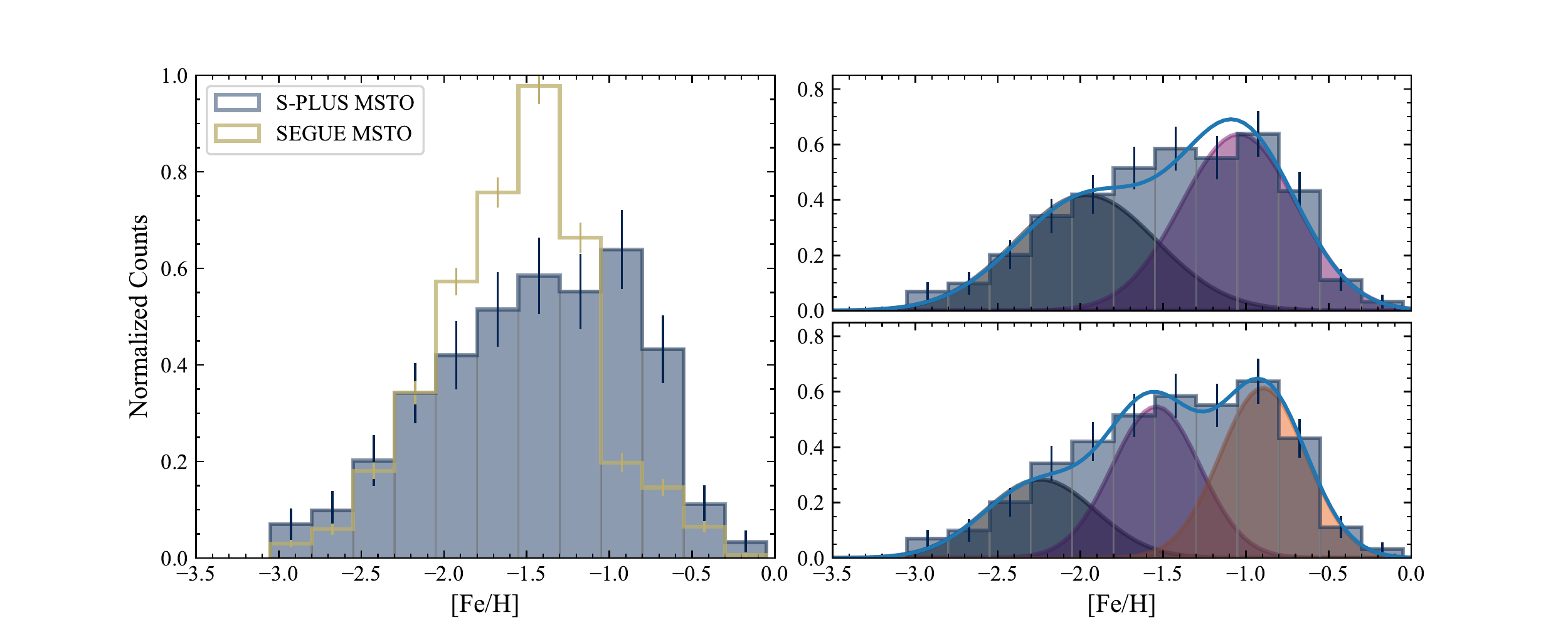}
	
	\caption{Metallicity distribution function of the S-PLUS and comparison SEGUE MSTO samples.  The S-PLUS MSTO sample (\textit{blue histogram}) photometric [Fe/H] distribution is shown in the left panel, along with the SEGUE spectroscopic distribution (\textit{tan open histogram}). The top right panel shows a two-component Gaussian Mixture Model produced for the S-PLUS sample is shown, indicated by the smooth dark blue and purple profiles. The blue line indicates the sum of the model distributions. A three-component Gaussian Mixture Model is shown in the bottom right panel.
	\label{fig:MSTO_FEH}}
\end{figure*}

\begin{figure}
	\centering
	\includegraphics[width=\columnwidth, trim=0.0cm 0.50cm 0.00cm 0.0cm, clip]{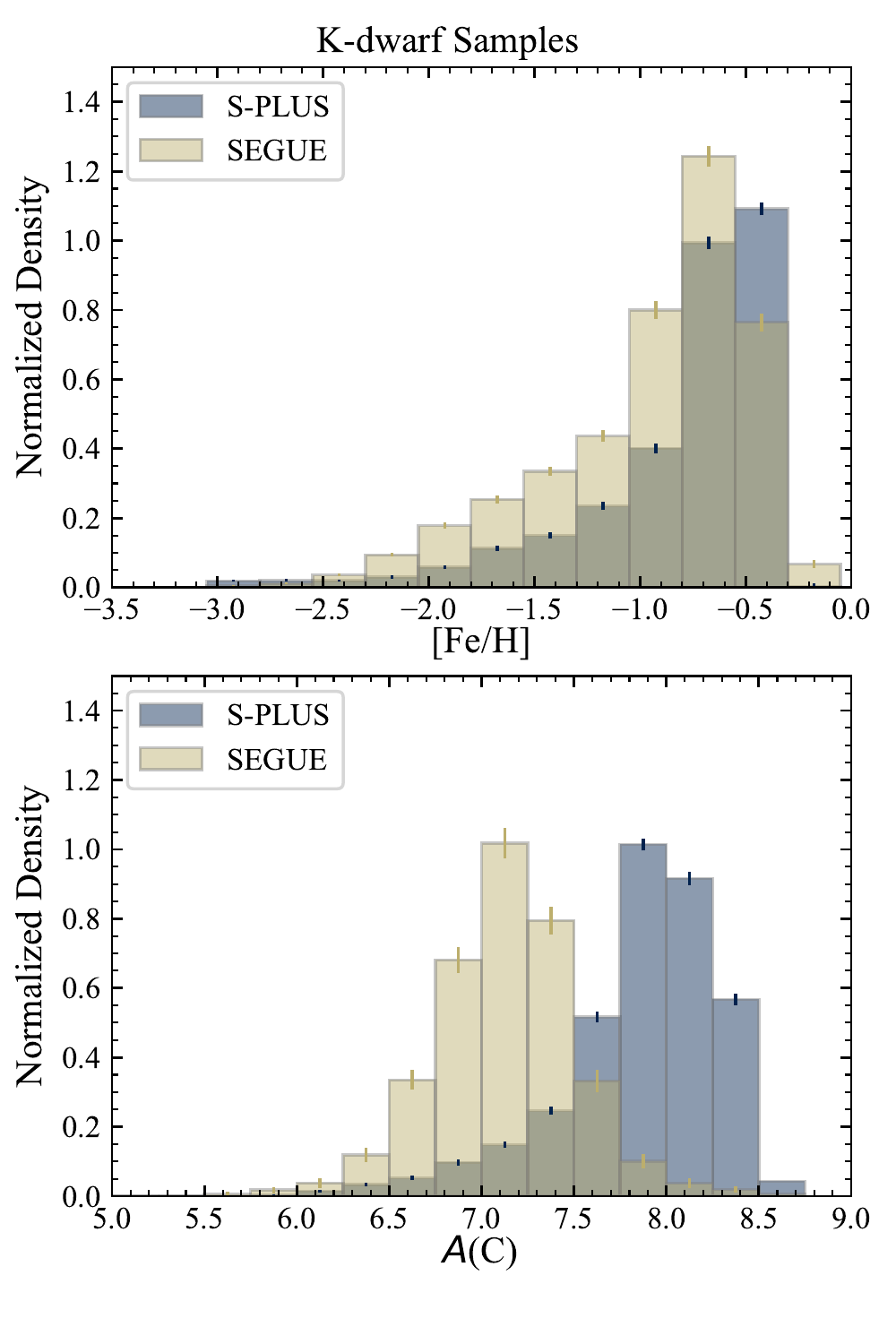}
	
	\caption{Metallicity and absolute carbon-abundance distributions of the S-PLUS and SEGUE K-dwarf samples. In the top panel, the [Fe/H] distribution of the S-PLUS sample (\textit{blue}) is shown along with the comparison SEGUE spectroscopic sample (\textit{tan}). Both samples are normalized to their peaks, [Fe/H]$=-0.64$ and [Fe/H]$=-0.81$ for the S-PLUS and SEGUE samples, respectively. The corresponding $A$(C) distribution is shown in the bottom panel. Both distributions are normalized to their peaks, $A({\rm C}) = 7.68$ and $A({\rm C}) = 7.13$ for the S-PLUS and SEGUE samples, respectively. \label{fig:dwarf_mdf}}
\end{figure}

Photometric estimates of effective temperature, metallicity, carbonicity, and absolute carbon abundance have been determined for 700,733 stars from the S-PLUS Stripe 82 Data Release 1 with revised calibration, using the Stellar Photometric Index Network Explorer (\texttt{SPHINX}) \citep{Whitten:2019a}. As described above, we made two independent selections, a MSTO selection for $6000 < T_{\rm eff} \textrm{(K)} < 7000$, and a mass-selected K-dwarf sample in the range $0.58 < M/M_{\odot} < 0.75$. 
The normalized MDFs and exponential slopes, $\lambda_{10}$, are provided in Table~\ref{table:MDF_table} for both samples, and we summarize their properties and implications below.

\subsection{MSTO Stars}\label{section:MSTO_results}

The MDF of the MSTO sample is shown in Figure~\ref{fig:MSTO_FEH}, along with the SEGUE spectroscopic sample for comparison. The distributions are limited to the magnitude range of $16.6 < g_0 <18.3$, corresponding to a distance range of $3 < d (\rm{kpc}) < 9$. Note that the apparent mismatch between the MSTO and SEGUE samples above [Fe/H] $= -2.0$ is expected, due to the metallicity bias in the selection of SEGUE targets, for which we have not corrected here.

The sample is clearly multi-modal, with at least two populations apparent. We address whether the MSTO MDF is better represented by two or three independent populations via application of GMM, indicated by the shaded PDFs in Figure~\ref{fig:MSTO_FEH}. While both models are plausible, a three-component model is determined to be superior on the basis of the Akaiki and Bayesian information criteria (AIC and BIC, respectively), where a difference between models of $\Delta \textrm{AIC}$ or $\Delta \textrm{BIC} > 10$ is indicative of a model with significantly higher loss of information \citep{Kass:1995}. 
Our three-component Gaussian model produces peaks at [Fe/H] = $-$0.9, $-$1.5, and $-$2.2. The two metal-poor components coincide with the inner and outer halos, respectively \citep{Chiba_2000, Carollo:2007,Carollo:2010,Beers:2012,dietz2020}. On the other hand, the metal-rich component is likely a combination of the canonical thick disk and the `splashed disk' \citep{bonaca2017,belokurov2020}, which has the same metallicity as the thick disk, but is extended farther away from the Galactic plane. This interpretation is also supported by the perspective of the local halo stars on the chemo-kinematical space in \citet{an2020,an2021}, according to which the local halo samples such as ours are dominated by the above three distinct stellar populations. Their mapping also suggests that the two Gaussian components centered at [Fe/H] = $-$0.9 and $-$1.5 are attributed to stars in the so-called red and blue sequences, respectively, in the Hertzsprung-Russell diagram obtained from Gaia DR2 (\citealt{gaia2018hrd}; see also \citealt{sahlholdt2019}).


Interestingly, we find that the outer-halo population component constitutes as much as 24\% of the stars in the MSTO sample, as close as 3\,kpc from the Sun. This is consistent with estimates produced by cosmological simulation studies (e.g., \citealt{Tissera:2014}), which suggest an outer-halo contribution of as much as $20 \mbox{--}40$\% in the Solar Neighborhood, based on hydrodynamical modeling \citep{Scannapieco:2009}. We describe the \citet{Tissera:2012,Tissera:2013,Tissera:2014} studies further in Section~\ref{section:aquarius}, in particular the role of accretions and \textit{in situ} star formation. 

\begin{table}
	\caption{Gaussian Mixture Model Components for MSTO Sample}             
	\label{table:gmm_parameters}      
	\centering                          
	\begin{tabular}{l c c c}        
		\hline\hline                  
		Component & $\mu$[Fe/H] & $\sigma$[Fe/H] & $\phi$\\  

		\hline   
        \multicolumn{4}{c}{Two-Component Model} \\
        \hline
        Comp 1   & $-1.04$ & $0.35$  & $0.55$ \\
        Comp 2   & $-1.96$ & $0.42$ &  $0.46$ \\
        \multicolumn{4}{c}{AIC = 8355, BIC = 8387} \\
        \hline
        \multicolumn{4}{c}{Three-Component Model} \\
        \hline
        Comp 1  & $-0.89$ & 0.26 & $0.40$ \\
        Comp 2  & $-1.54$ & 0.26 & $0.36$ \\
        Comp 3  & $-2.24$ & 0.33 & $0.24$ \\
        \multicolumn{4}{c}{AIC = 8261, BIC = 8312} \\
        \hline        
		\hline                                   
	\end{tabular}
\end{table}


\begin{table}
	\caption{MDFs for MSTO and K-dwarf Stars}             
	\label{table:MDF_table}      
	\centering                          
	\begin{tabular}{c c c r c}        
		\hline\hline                  
		& MSTO & K-dwarf & MSTO & K-dwarf \\
		 \textrm{[Fe/H]} & $N/N_{\rm tot}$   &  $N/N_{\rm tot}$   &    $\lambda_{10} $    &  $\lambda_{10}$ \\ 
		\hline
		$-1.0$      &  $0.157$  &  $ 0.323 $   &  $  0.00 $ &  $ 0.91 $      \\
		$-1.2$      &  $0.139$  &  $ 0.214 $   &  $  0.09 $ &  $ 0.85 $        \\
		$-1.4$      &  $0.145$  &  $ 0.146 $   &  $  0.02 $ &  $ 0.76 $      \\
		$-1.6$      &  $0.125$  &  $ 0.107 $    & $  0.22 $ &  $ 0.73 $       \\
		$-1.8$      &  $0.123$  &  $ 0.074 $    & $  0.32 $ &  $ 0.91 $       \\
		$-2.0$      &  $0.097$  &  $ 0.043 $    & $  0.51 $ &  $ 1.06 $        \\
		$-2.2$      &  $0.078$  &  $ 0.028 $    & $  0.40 $ &  $ 0.88 $       \\
		$-2.4$      &  $0.061$  &  $ 0.020 $    & $  1.14 $ &  $ 0.55 $        \\
		$-2.6$      &  $0.031$  &  $ 0.016 $    & $  1.04 $ &  $ 0.38 $    \\
		$-2.8$      &  $0.029$  &  $ 0.014 $    & $  0.80 $ &  $ 0.59 $    \\
		$-3.0$      &  $0.010$  &  $ 0.009 $    & $  2.82 $ &  $ 1.27 $    \\
		\hline        
		\hline
	\end{tabular}
\end{table}

\subsection{Mass-Selected Local K-Dwarfs}

To select for local K-dwarf stars in the mass range $0.58 < M/M_{\odot} < 0.75$, we implemented a Support Vector Classification (SVC) scheme using the combination of the broad-band S-PLUS color $(g-i)_0$ and estimates of effective temperature and metallicity obtained from \texttt{SPHINX}. The SVC mass selection was performed on the S-PLUS Stripe 82 sample, with a subsequent RFC giant rejection (see Section~\ref{sec:dwarf_selection} above), resulting in 52,035 stars with robustly determined metallicity and absolute carbon-abundance estimates within the heliocentric distance $d<6$\,kpc. As this selection constituted a Solar Neighborhood sample, no minimum distance was implemented. A similar selection is made for our SEGUE spectroscopic sample, resulting in 12,337 stars, over heliocentric distances of $1 < d \textrm{(kpc)} < 6$. 

The resulting metallicity and absolute carbon-abundance distributions are shown in Figure~\ref{fig:dwarf_mdf}. The peaks of the [Fe/H] and $A$(C) distributions are clearly distinct for S-PLUS and SEGUE. The S-PLUS K-dwarf MDF peaks at a higher metallicity, [Fe/H] $=-0.64$, compared to the SEGUE value of [Fe/H] $=-0.81$. Similarly, the $A$(C) distribution of the S-PLUS sample peaks at $A({\rm C})=7.85$, compared to $A({\rm C})=7.13$ for the SEGUE sample. The largest influence on this separation is likely the spectroscopic selection function of SEGUE, which is (by design) biased towards low-metallicity stars. For stars of a given [C/Fe], a lower peak [Fe/H] would result in a lower peak $A$(C), as it is also influenced by the previously known relationship between increasing [C/Fe] with decreasing [Fe/H] (see, e.g., \citealt{Yoon:2018:AEGIS}, and references therein).  However, the difference between the peak of the [Fe/H] distributions between the S-PLUS and SEGUE K-dwarf sample of $0.23$\,dex is not sufficient to account for the $A$(C) peak separation of $0.54$\,dex. Given that the K-dwarf selection procedure was implemented identically for both the S-PLUS and SEGUE samples, we conclude that an additional characteristic of the SEGUE selection function is influencing the $A$(C) distribution, other than the known bias towards low-metallicity stars.


\begin{figure}
	\centering
	\includegraphics[width=\columnwidth, trim=0.0cm 0.50cm 0.50cm 0.00cm, clip]{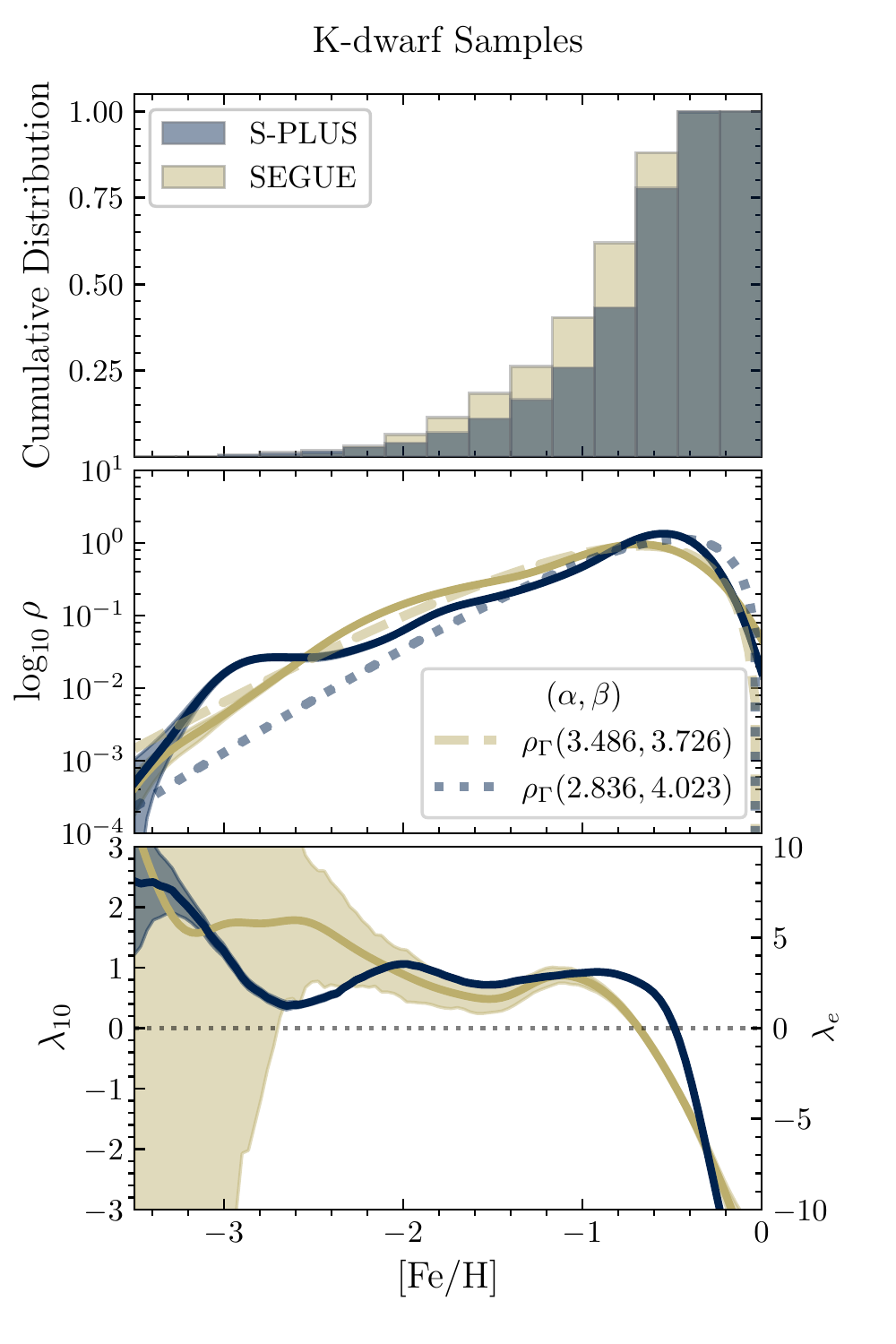}
	
	\caption{Empirical metallicity distributions of the mass-selected K-dwarf samples from S-PLUS Stripe 82 and SEGUE. In the top panel, the normalized cumulative metallicity distribution is shown for S-PLUS (blue) and SEGUE (tan). The logarithm of the probability distribution is shown in the middle panel, as determined by kernel-density estimation. The gradient of the MDF is shown in the bottom panel as a function of metallicity, with $\lambda_{10}$ and $\lambda_e$ given on the left and right of the y-axes, respectively. Uncertainties are determined via successive bootstrap resampling.
		\label{fig:dwarf_mdf_lambda}}
\end{figure}


\subsubsection{Comparison of the K-dwarf Metallicity Distributions}

We further explore the shapes of the S-PLUS and SEGUE K-dwarf Stripe 82 MDF in Figure~\ref{fig:dwarf_mdf_lambda}. The cumulative distributions of these distributions is shown in the top panel, with the MDF in $\log_{10}(\rho)$ given in the center panel. The value of the exponential slope, as defined in Eqn.~\ref{eq:expo_e} and Eqn.~\ref{eq:expo_10}, is shown as a function of the metallicity in the bottom panel.

The S-PLUS MDF exhibits a steeper profile than the SEGUE profile within the metallicity range $-2.00 < \textrm{[Fe/H]} < -1.25$, $\lambda_{10} = 0.91\pm0.02$ for the S-PLUS sample and $\lambda_{10} = 0.70\pm0.02$ for the SEGUE sample. The SEGUE MDF continues to decline steadily below [Fe/H] $<-2.0$, while the S-PLUS MDF levels off until [Fe/H] $<-3.0$, beyond which it falls rapidly. The behavior of the exponential slope is depicted in the bottom panel of Figure~\ref{fig:dwarf_mdf_lambda}, where the instantaneous value of the slope, $\lambda_{10}$, is determined as a function of metallicity. 

 The SEGUE sample exhibits substantially larger scatter in all parameter estimates for both distributions, most notably the exponential slope of $\sigma(\lambda_{e}) = 1.06$. These values were determined considering the metallicity range of [Fe/H] = [$-2.5$, $-0.5$].  Figure~\ref{fig:dwarf_mdf_lambda} shows, in the top panel, the cumulative frequency distributions of [Fe/H] for the two samples, and in the center panel, the $\log_{10}(\rho)$ of the MDFs. The MDFs in this panel are determined via kernel-density estimation, averaged over 300 successive bootstrap resamples. To further explore the dependency of the exponential slope on metallicity, we compute the numerical gradient of the MDFs for each sample, shown in the bottom panel.  The gradient of $\log_{10}(\rho)$ is seen to vary with metallicity. We note again the significantly smaller uncertainties of $\lambda_{\rm 10}$ for the S-PLUS sample, represented by the shading in this panel, where this uncertainty becomes overwhelming for the SEGUE spectroscopic sample below [Fe/H] $=-2.5$.
Not surprisingly, a one-tail Kolmogorov-Smirnov 
test rejects the null hypothesis that both samples are drawn from the same parent population ($p = 0.01$).

\begin{table*}
	\caption{Maximum Likelihood Parameter Estimates for K-dwarf Sample Metallicity Distributions}             
	\label{table:parameters}      
	\centering                          
	\begin{tabular}{l c c c c c c}        
		\hline\hline                  
 
        \multicolumn{7}{c}{Exponential Distribution ($\rho_{e}$)} \\
        Sample & $\lambda_{10}$ & $\sigma(\lambda_{10})$ & $\lambda_e$ & $\sigma(\lambda_e)$  & AIC & BIC \\
        \hline
        S-PLUS              & $0.85$   & $0.26$   & $1.95$    &  $ 0.59$  &   $31144$  & $31153$  \\
        SEGUE                 & $0.70$   & $0.46$   & $1.62$   &  $1.06$   &   $14559$  & $14566$ \\
        \hline
        
        \multicolumn{7}{c}{Gamma Distribution ($\rho_{\Gamma}$)} \\
        Sample & $\alpha$ & $\sigma(\alpha)$ & $\beta$ & $\sigma(\beta)$  & AIC & BIC \\
        \hline
        S-PLUS               &    $2.84$     &   $0.02$    &  $4.02$    & $0.03$   & $\mathbf{28505}$  &  $28523$   \\
        SEGUE                 &    $3.49$    &   $0.08$     &  $3.73$     & $0.10$  &  $\mathbf{12577}$  &  $12591$     \\
        \hline        

		\hline                                   
	\end{tabular}

\end{table*}

We model the K-dwarf sample MDF with a gamma function, Eqn.~\ref{eq:beta_pdf}, using MLE. The resulting posterior distributions of the $\alpha$ and $\beta$ are listed in Table~\ref{table:parameters}. The uncertainties are reported as the standard deviation of the posterior distributions, and are seen to again be considerably smaller for the S-PLUS sample. This is likely more reflective of the gamma distribution's flexibility and overall ease of determination. To build more representative uncertainties in the most likely values of $\alpha$ and $\beta$, we iteratively resample both distributions with bootstrap and successive MCMC determinations. Both samples exhibit similar $\beta$ values, $\beta = 4.02 \pm 0.03$ and $\beta= 3.73 \pm 0.10$ for the S-PLUS and SEGUE samples, respectively. As this is a two-parameter distribution, and the mean metallicity of both samples is similar, this is not remarkable. However, the optimal values of the shape parameter are markedly different: $\alpha = 2.84 \pm 0.02$ for the S-PLUS sample, and $\alpha=3.49 \pm 0.08$ for the SEGUE sample. The resulting best-fit gamma distributions are shown in Figure~\ref{fig:dwarf_mdf_lambda}.

The gamma distribution is seen to be a more appropriate representation of the metallicity distribution of our K-dwarf sample than an exponential, evidenced by the superior Akaiki and Bayesian Information Criteria in Table~\ref{table:parameters}. The AIC and BIC weigh the maximized likelihood against the model's degrees of freedom. Models with a smaller AIC or BIC are preferred, where $\Delta \textrm{BIC} > 10$ is considered significant. These values correspond to Fisher-Pearson coefficients of skewness of $-1.19 \pm 0.01$ and $-1.0 \pm 0.1$ for the S-PLUS and SEGUE MDFs, respectively. 

While both the S-PLUS and SEGUE MDF slopes are consistent with previous studies, we note that the value of $\lambda_{10}$ varies significantly across the full metallicity range considered. It is therefore of particular importance to consider the metallicity across which the exponential slope is being considered. At the low metallicity range, $-3.4 < \textrm{[Fe/H]} < -2.5$, \citet{Youakim:2020} reports a slope of $1.0 \pm 0.1$\footnote{The value reported as $\Delta(\log{\rm N})/\Delta\textrm{[Fe/H]}$ in the \citet{Youakim:2020} study is equivalent to the $\lambda_{10}$ used in this work.}. We find a similar value at a higher metallicity range, $-1.8 < \textrm{[Fe/H]} < -1.0$, but find an essentially flat slope around [Fe/H] $=-2.5$, $\lambda_{10} = 0.1$. Below [Fe/H]  $<-2.75$, the slope increases sharply, reaching $\lambda_{10} = 2.11$ at [Fe/H]  $=-3.0$. This is larger than the slope derived by \citet{DaCosta:2019} of $1.5 \pm 0.1$, who studied a spectroscopic sample of metal-poor giants selected from DR 1.1 of the SkyMapper survey. Nevertheless, our result supports the conclusion that the MDF drops more steeply at the lowest metallicity. We do not attribute the rapid increase in the slope seen in Figure~\ref{fig:dwarf_mdf_lambda} to the size of our sample. Below [Fe/H] $<-2.75$, we have 467 candidates, with 61 stars below [Fe/H] $<-3.0$.

\begin{figure*}
	\centering
	\includegraphics[width=\textwidth, trim=1.50cm 0.75cm 1.750cm 0.5cm, clip]{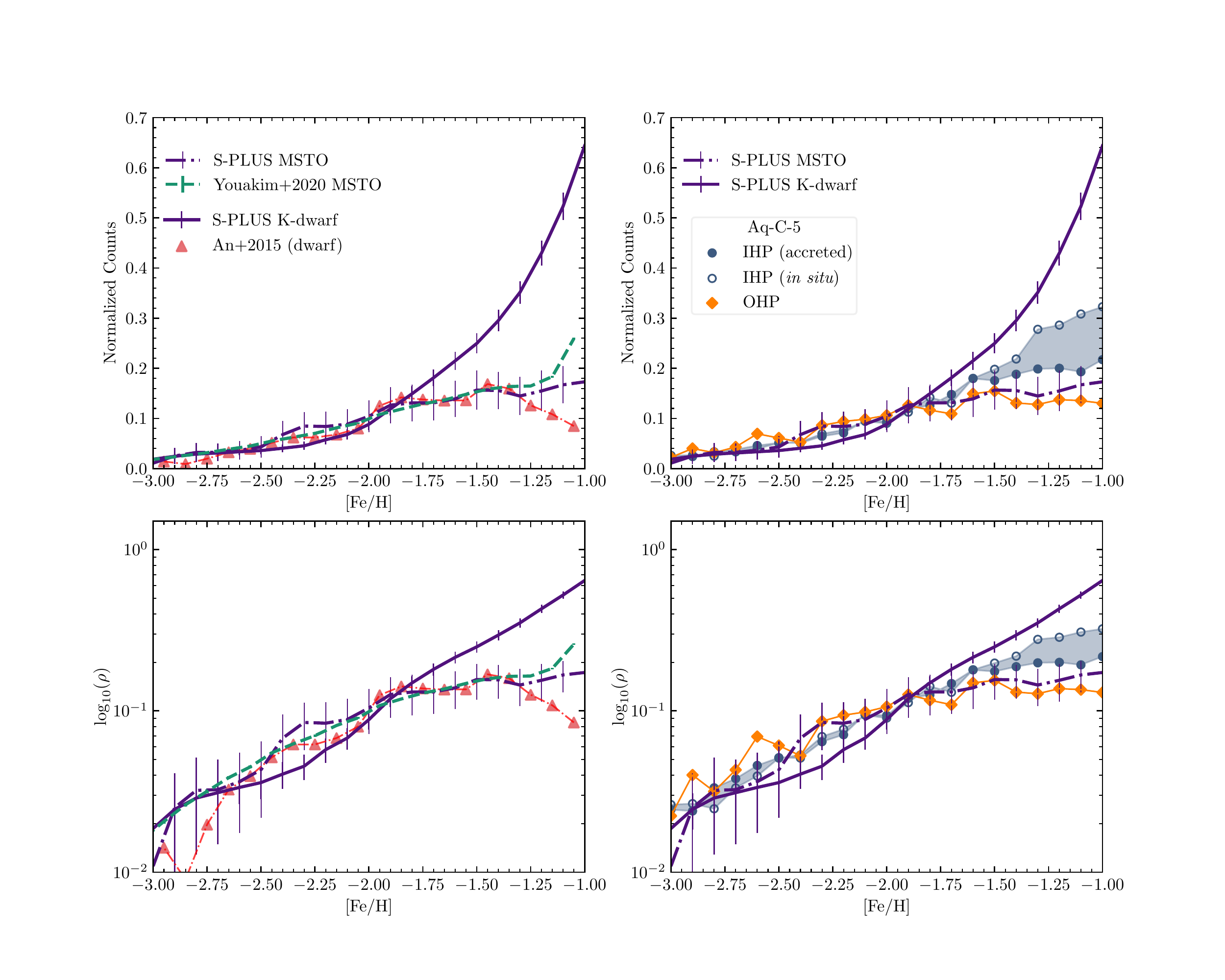}
	\caption{Metallicity distribution functions of the S-PLUS MSTO (\textit{purple, dotted}) and K-dwarf (\textit{purple, solid}) samples compared with literature studies. In the left panel, the observational studies from \citet{An:2015} (\textit{red, triangles}) and \citet{Youakim:2020} (\textit{green, dashed}) are shown in comparison to the S-PLUS samples. (\textit{Right:}) Simulation results for the inner-halo population (IHP; \textit{dark-blue circles}) and outer-halo population (OHP; \textit{orange diamonds}) of the \texttt{Aq-C-5} \citet{Tissera:2014} halo sample are shown for comparison. The IHP of \texttt{Aq-C-5} is further divided into the accreted (\textit{filled points}) and \textit{in situ} (\textit{hollow points}) components. The shading indicates the region where we expect MDFs composed of contributions from both components (see text for details). The logarithm of each distribution is shown in the bottom panels. All distributions are normalized to the interval [Fe/H] $=[-2.0,-1.5]$ for comparison.\label{fig:MDF_Compare}}
\end{figure*}

\subsection{Comparison with Previous Work}

Figure~\ref{fig:MDF_Compare} shows a comparison of our MDF results with those from the studies of \citet{An:2015} and \citet{Youakim:2020}, and considered in more detail below.

\subsubsection{Comparison with \citet{An:2013,An:2015}}

As described in \S~4.3, we made a mass-based selection for the K-dwarf
sample for a number of reasons. One of the advantages of such a selection
is that we can reduce bias against metallicity, which is normally
inherent to color- or $T_{\rm eff}$-based selections due to a strong
correlation between metallicity and $T_{\rm eff}$. \citet{An:2013,An:2015}
also made such attempts to construct \textit{in situ} halo samples of
main-sequence dwarfs in Stripe~82. Their selection, based on stellar
masses in the range $0.65 < M/M_\odot < 0.75$, includes approximately
$2,500$ sources in each of three photometric catalogs, which have been
independently generated from multiple SDSS imaging data.

The top panels of Figure~\ref{fig:MDF_Compare} show a comparison of the normalized
MDFs from the S-PLUS K-dwarf sample (purple
solid line) with that in \citet[][red triangles]{An:2015}. While the
S-PLUS K-dwarf sample is steeply rising towards higher metallicities,
the \citet{An:2015} MDF levels off at [Fe/H]~$\sim-1.5$. The large
discrepancy at higher metallicities originates from important
differences in the detailed sample selections in these two studies. The
\citet{An:2013,An:2015} samples are strictly limited to sources at $5 < d
\la10$~kpc with $|b| > 35\arcdeg$. On the other hand, the S-PLUS sample
includes more local stars in the thick disk ($d > 1$~kpc), resulting in
a factor of ten larger sample size, and more importantly, leading to a
skewed MDF by including significantly more metal-rich disk (and
metal-weak thick disk) stars.

Nonetheless, at [Fe/H]~$<-1.7$, where the contributions from metal-rich
disk stars are minimal, the observed shape of the MDF from the S-PLUS
K-dwarf sample is similar to that of \citet{An:2015}, considering the size
of the errors from both studies. Interestingly, the latter MDF agrees better
with that of the S-PLUS MSTO sample (purple dotted line), at least up to
[Fe/H]~$\sim-1.3$. \citet{An:2013,An:2015} judged that their MDF is complete
at [Fe/H]~$<-1.2$, above which the number of metal-rich stars are
under-estimated at large distances, because more metal-rich stars in a
given mass bin become fainter, and therefore are under-populated in a
magnitude-limited sample. The metallicity bias for the metal-poor star
samples were mitigated in their studies by adopting detailed
mass-metallicity-distance cuts.\footnote{The same sample cut was not
applied to the S-PLUS K-dwarf sample in this exploratory study, in order
to maximize the number of metal-poor stars in the carbonicity estimation
(see below).} Therefore, the agreement between the MSTO sample and
\citet{An:2015} at [Fe/H]~$\la-1.2$ is reassuring, given the fact that
they are based on different imaging data and independent metallicity
estimators.

When decomposed into a two-component Gaussian model, \citet{An:2013,
An:2015} found peaks at [Fe/H]~$=-1.5\pm 0.2$ and [Fe/H] ~$=-2.1\pm0.2$ for
the inner- and outer-halo components, respectively, taken as an average
from the three photometric catalogs employed in their studies. These
peak values are in good agreement with the peaks identified from the
three-component Gaussian decomposition of our MSTO sample (Table~\ref{table:MDF_table}); the
metal-rich component centered at [Fe/H]~$= -0.9$ can be attributed to the
thick-disk and metal-weak thick disk stars in the MSTO sample with $d > 3$~kpc, which are not
present in \citet{An:2013,An:2015}. We also find that the fraction of the
low-metallicity halo component from the MSTO sample ($24\%$) is
compatible with the $\sim20$--$35\%$ in \citet{An:2013}, although it is smaller
than the $\sim35$--$55\%$ reported in \citet{An:2015}.

\subsubsection{Comparison with \citet{Youakim:2020}}

We now compare our MSTO sample with the \citet{Youakim:2020} study of MSTO stars from the {\it Pristine} survey \citep{Starkenburg:2017}. Note that the \citet{Youakim:2020} selection covered a larger footprint than our S-PLUS sample, some $2,500$\,deg$^2$, predominately above the Galactic plane, with $|b| > 20^{\circ}$, although it included SDSS Stripe 82. Their selection also extends to larger heliocentric distances,  $6 < d~(\textrm{kpc}) < 20$, as compared with our MSTO sample range of $3 < d~(\textrm{kpc}) < 9$. Similar to \cite{An:2015}, the \citet{Youakim:2020} photometric methodology becomes unreliable for metallicities above [Fe/H] $ > -1.5$. For our MSTO sample, we do not expect such bias until [Fe/H] $>-0.75$. 

Similar to our MSTO selection, the \textit{Pristine} selection suggests the presence of three populations, with peak metallicities corresponding to [Fe/H] $=-1.32, -1.89$, and $-2.45$. The most metal-poor component is in general agreement with the value determined for our MSTO selection (see Table 1), [Fe/H] $=-2.51$, although our fraction of this population, 27\%, is somewhat smaller than determined for the \textit{Pristine} sample, 33\%. This is likely due to the larger contribution from thick-disk system stars to our MSTO sample, based on $ d > 3$\,kpc. 

\subsection{Comparison with the Aquarius Simulation Aq-C-5 of Tissera et al.} \label{section:aquarius}

Over the past decade, cosmologically based numerical simulations of the formation and evolution of Milky-Way-size systems have grown in sophistication to the point that insight into the chemical-evolution history of our galaxy can be gained by comparison with observations such as ours.  

Among these simulation studies is the Aquarius project, originally presented in \citet{Scannapieco:2009}. This high-resolution hydrodynamical simulation was produced with a modified version of GADGET-3 \citep{Scannapieco:2005, Scannapieco:2006} as part of the Aquarius Project of the Virgo Consortium \citep{Springel:2008}, and explicitly models the chemical evolution of the baryons via contributions from Type II and Type Ia supernovae. 
We refer the interested reader to \citet{Scannapieco:2009} for a detailed discussion of the chemical modeling, and focus our attention on one of the MW-mass halos studied by \citet{Tissera:2013,Tissera:2014}. In \citet{Tissera:2013}, a simple binding-energy criteria was developed to separate the halo component of the Aquarius MW-mass galaxies into an inner-halo population (IHP) and and outer-halo population (OHP). By tracking contributions of stars formed both within and outside of the main progenitor's virial radius, \citet{Tissera:2014} highlighted the contrasting roles of accretion and \textit{in situ} star formation in the evolution of MW-mass halos, and most importantly, the influence of these two mechanisms on the resulting chemo-dynamical distributions. 

We compare our observations with a particular halo from Aquarius, referred to as \texttt{Aq-C-5}. Of the eight MW-mass halos produced in Aquarius, \texttt{Aq-C-5} was determined to be the most similar to the MW halo population, on the basis of metallicity and age-distribution studies from \citet{Carollo:2018} and \citet{Whitten:2019b}, the chemo-dynamical properties of the old population in the central regions \citep{Tissera:2018}, and the [$\alpha$/Fe] radial profiles in the inner region \citep{Fernandez-Alvar:2019}. We restrict our analysis to the portion of the stellar halo of \texttt{Aq-C-5} with heliocentric distances in the range $3 < d \rm{(kpc)} < 9$. MDFs determined for the \texttt{Aq-C-5} halo were weighted according to the particle mass, as these particles represent conglomerates of stars. In the right panels of Figure~\ref{fig:MDF_Compare}, we consider the MDFs of \texttt{Aq-C-5} halo, which includes the OHP, and \textit{in situ} and accreted components of the IHP. Here, the IHP is indicated by light-blue circles; the OHP is shown in orange. Shading indicates the transition between the accreted and \textit{in situ} components of the IHP, or in other words, the region we expect to be occupied by MDFs composed of contributions from the two extremes of halo formation, namely pure accretion or pure \textit{in situ} star formation.

Across the metallicity range $-2.5 < \textrm{[Fe/H]} < -1.5$, the MDF of the S-PLUS MSTO sample is consistent with the \texttt{Aq-C-5} IHPs. Above [Fe/H] = $-1.50$, the MDF of the S-PLUS MSTO sample is more consistent with the OHP, although we acknowledge the limitations of this comparison, and do not suggest that the MSTO sample is dominated by outer-halo stars at [Fe/H] $>-1.50$. Below [Fe/H] $= -2.5$, the increase in stars noted in the S-PLUS MSTO MDF is also seen in the MDF of the OHP of \texttt{Aq-C-5}.

Similar to the S-PLUS MSTO sample, the S-PLUS K-dwarf sample MDF is consistent with the IHP of \texttt{Aq-C-5} within the range $-2.5 < \textrm{[Fe/H]} < -1.5$. At [Fe/H] $> -1.5$, however, the S-PLUS K-dwarf sample greatly exceeds both the inner- and outer-halo populations of \texttt{Aq-C-5}. We note that the \texttt{Aq-C-5} stellar halos do not include stars with circularity ($J_z/J_{z,\textrm{max}(E)}$) larger than 0.65 \citep[see][for details]{Tissera:2013}, effectively removing the disk population that is known to dominate the S-PLUS K-dwarf sample. This limits the extent to which comparisons between the S-PLUS K-dwarf sample and \texttt{Aq-C-5} can be made at higher metallicities, where the disk system is expected to dominate. Below [Fe/H] = $-2.0$, however, this complication is minimal, and we see that the K-dwarf MDF is somewhat steeper than both the inner- and outer-halo populations.

It is clear that future refinements of the photometric MDFs, such as ours, can help constrain the stellar contributions from the accreted and {\it in situ} populations predicted by such simulations. Interestingly, it is the regime of the MDF at higher-metallicity, [Fe/H] $>-1.50$, which exhibits the largest contrast between accreted and \textit{in situ} populations, and is thus of particular importance to constrain their respective roles.\\

\subsection{Photometric Absolute Carbon Abundance}

\begin{figure}
	\centering
	\includegraphics[width=\columnwidth, trim=0.0cm 0.00cm 3.00cm 1.50cm, clip]{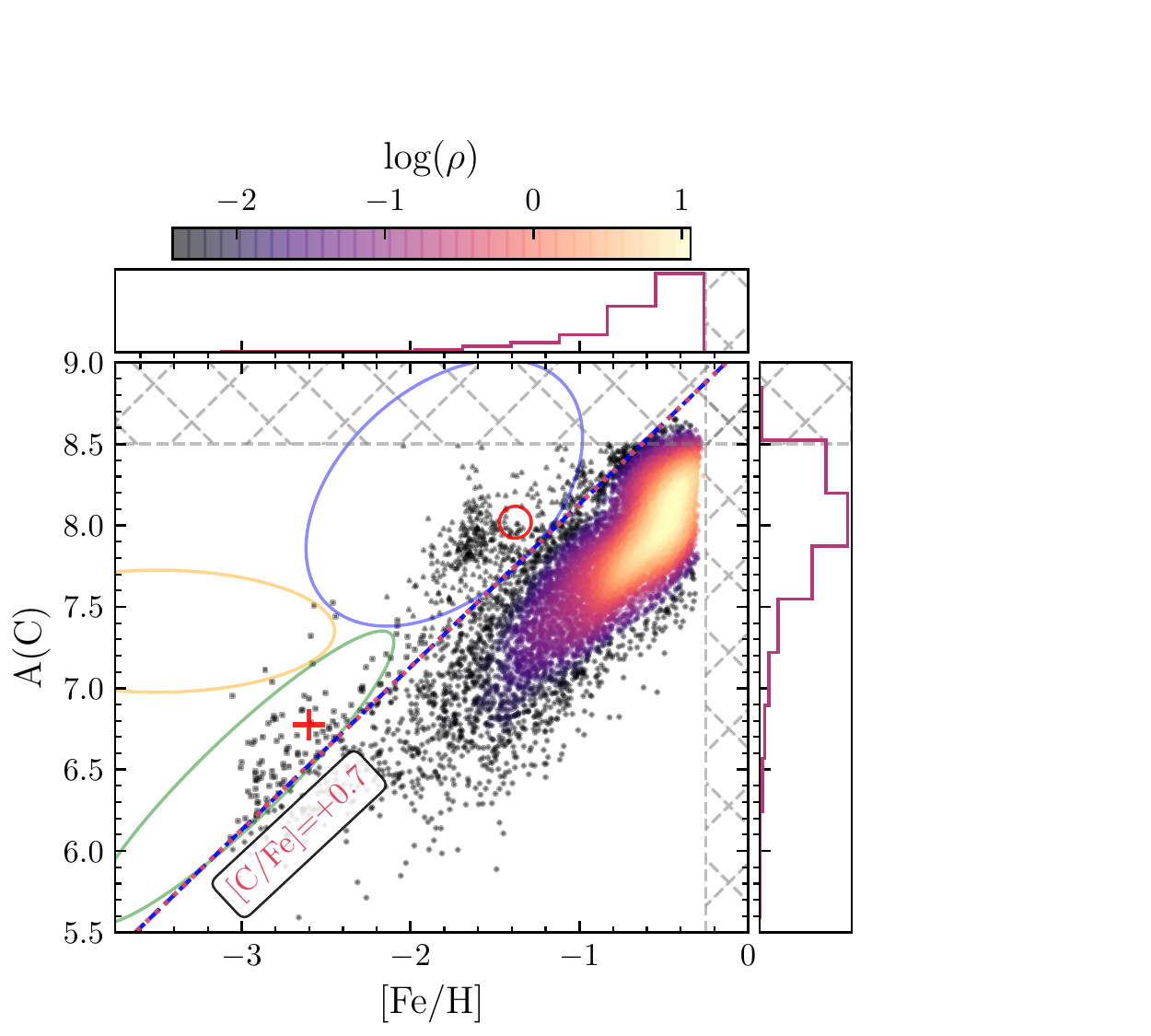}
	
	\caption{[Fe/H] vs. $A$(C) distribution of the S-PLUS Stripe 82 K-dwarf sample. The number density ($\rho$) is determined by kernel-density estimation and represented by the color-coding shown. The line corresponding to [C/Fe] $=+0.7$ indicates the adopted boundary for carbon-enhanced stars, with two populations clearly present in the CEMP region. The red cross and circle markers at ([Fe/H], $A$(C)) $= (-2.8, 6.5), (-1.6, 8.0)$ represent the cluster centers determined by K-means clustering. Ellipses corresponding approximately to the CEMP Group I (\textit{blue}), II (\textit{green}), and III (\textit{orange}) identified in \citet{Yoon:2016} are shown for reference. \label{fig:YOON_BEERS}}
\end{figure}

Estimates of absolute carbon abundance, $A$(C) (and [C/Fe]), are obtained for the S-PLUS K-dwarf sample, resulting in robust determinations for $51,003$ stars across the carbon abundance range $5.0 < A({\rm C}) < 8.5$. Figure~\ref{fig:YOON_BEERS} shows $A$(C) vs. [Fe/H], following \citet{Yoon:2016}. Here, the color-coding indicates the logarithmic number density, determined from a 2-D kernel-density estimate with a standard deviation assigned to the typical estimated uncertainty in $A$(C), $\sim 0.35$\,dex. The dashed line corresponds to [C/Fe] $= +0.7$, the level of carbonicity used to separate CEMP stars from carbon-normal stars. 
This diagram also indicates the approximate location of the morphological Group I, Group II, and Group III stars suggested by \citet{Yoon:2016}, although our inability to estimate [Fe/H] for stars below [Fe/H] $ = -3.5$ limits the number of lower-metallicity Group II and Group III stars we can identify. 


A total of 364 stars populate the CEMP region above [C/Fe]  $=+0.7$. 
Excepting the small number of possible Group III stars, two distinct clusters are apparent among the CEMP stars, roughly consistent with the morphology identified by \citet{Yoon:2016}, namely a large grouping of stars $-2.5 <  {\rm [Fe/H]} <  -1.0$ and $7.5 <  A{\rm (C)}  < 8.5$, and a separate population at $-3.5 <  {\rm [Fe/H]} <  -2.0$ and  $6.0 < $ $A$(C) $ < 7.25$. We model these clusters using a K-means approach, for which we determine cluster centers at $(\textrm{[Fe/H]}, A\textrm{(C)}) = (-1.6, 8.0)$ and ([Fe/H], $A\textrm{(C)}) = (-2.8, 6.5)$, respectively. These clusters are represented in Figure~\ref{fig:YOON_BEERS} by the red circle and cross, respectively. The stars associated with the higher-metallicity cluster constitute $\sim 70$\,\% of the CEMP stars in our dwarf sample.

\citet{Yoon:2016} identified similar populations in $A$(C) vs. [Fe/H], produced from  a compilation of high-resolution spectroscopic determinations, which was built upon the literature compilation of \citet{Placco:2014}. The peak at $A(\rm{C})\sim 8.0$ in \citet{Yoon:2016}, designated as Group I, was found to constitute the majority of CEMP-$s$ and CEMP-$r/s$ stars\footnote{CEMP-$s$: [C/Fe] $>+0.7$, [Ba/Fe] $ > +1.0$, and [Ba/Eu] $>+0.5$ \newline CEMP-$r/s$: [C/Fe] $ > +0.7$ and $0.0 < $ [Ba/Eu] $ <+0.5$ \newline CEMP-no : [C/Fe] $ > +0.7$ and [Ba/Fe] $ < 0.0$} stars, while a lower peak at $A(\rm{C}) \sim 6.3$, designated Group II, was found to possess a majority of CEMP-no stars. This diversification is a compelling distinction, with implications on the first-generation star progenitors required to produce the observed CEMP star behavior. We note that a third population, Group III, is located at lower metallicity, [Fe/H] $<-3.0$, and centered on $A(\rm{C})\sim 6.8$. Group III CEMP stars are also predominantly CEMP-no stars, and are perhaps the most interesting with regards to the impact on the origin of carbon. However, as noted above, Group III stars with [Fe/H] $<-3.5$ are effectively inaccessible with the current limitations of the photometric metallicity technique central to this work. 

For these reasons, we identify the clusters identified by our K-means method as the Group I and Group II populations discussed in \citet{Yoon:2016}, and suggest that the handful of stars seen in Figure~\ref{fig:YOON_BEERS} at lower metallicities, [Fe/H]$<-2.5$, and $A(\rm{C}) \sim 7.0-7.5$ may be Group III CEMP stars in the S-PLUS K-dwarf sample.  

\begin{figure*}
	\centering
	\includegraphics[width=\textwidth, trim=2.20cm 0.00cm 2.30cm 0.00cm, clip]{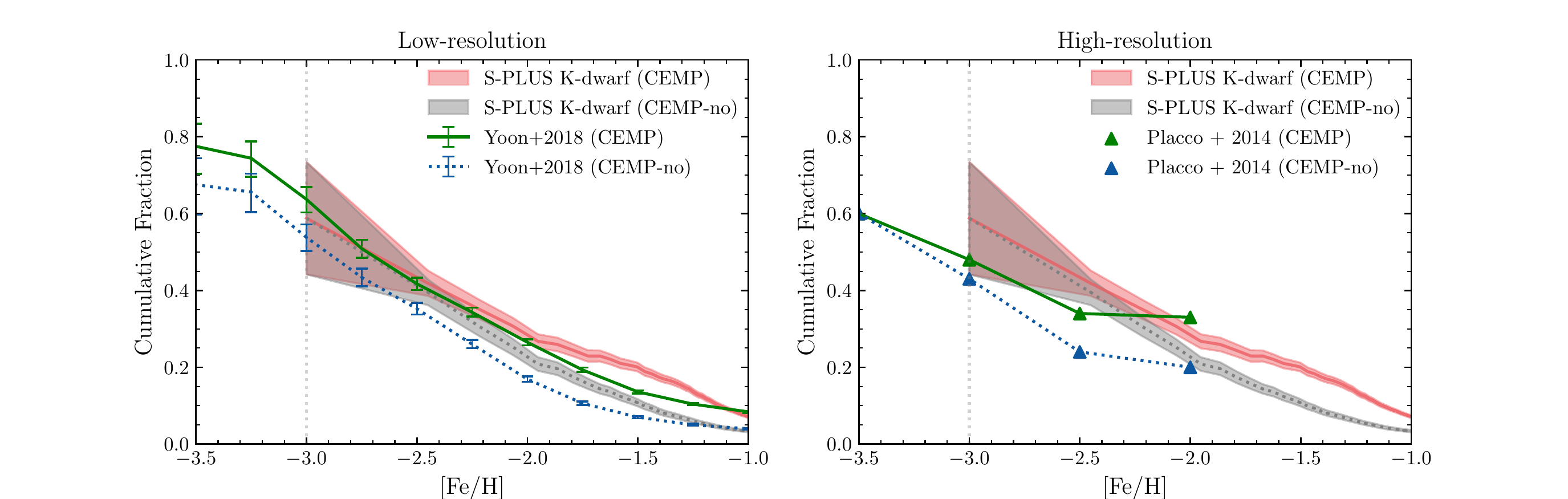}
	
	\caption{Cumulative CEMP-star fractions for the S-PLUS K-dwarf sample compared with literature studies.  In the left panel, the S-PLUS K-dwarf cumulative CEMP-star fraction  (\textit{red, shaded}) and CEMP-no star fraction is shown along with the \citet{Yoon:2018:AEGIS} study produced with medium-resolution spectroscopy from AEGIS. The \citet{Placco:2014} study produced with high-resolution spectroscopy is shown in the right panel. Shading on the K-dwarf CEMP and CEMP-no frequencies represents the binomial confidence intervals, averaged by successive bootstraping. The dotted vertical line represents the current limit of this work, below which the number of stars is insufficient.\label{fig:CEMP_frac}}. 
\end{figure*}

\subsubsection{Cumulative CEMP Fraction Comparison}

The overall frequency of CEMP stars in our K-dwarf sample is 1.6\%, and increases rapidly at lower metallicity. Nearly half, 48\,\%, of all stars in our sample below [Fe/H] $ <-2.5$ are CEMP stars. Of the 14 stars below [Fe/H] $< -3.0$, 14 (78\%) are CEMP stars.
We compare this result with with two studies from the literature, \citet{Yoon:2018:AEGIS} and \citet{Placco:2014}. First, the \citet{Yoon:2018:AEGIS} study explored the spatial distributions of [C/Fe] and [Fe/H] with medium-resolution ($R \sim 1,300$) spectroscopy from the AAOmega Evolution of Galactic Structure; AEGIS) survey. This study conveniently includes regions of the Southern Galactic Hemisphere, permitting comparisons with Stripe 82. 
However, the \citet{Yoon:2018:AEGIS} study was concerned with sub-giants and giants, thus probing the CEMP fraction at considerably larger heliocentric distances than the S-PLUS K-dwarf sample.
The \citet{Placco:2014} study obtains the CEMP fraction based on a compilation of the high-resolution spectroscopy available at the time, from the SAGA database \citep{Suda:2008} and the literature compilation of \citet{Frebel:2010}. This sample also includes a large fraction of giants, as well as stars with high effective temperatures ($T_{\rm{eff}} > 6000$~K), and thus a similar caution is given with regards to the extent with which comparisons can be made. We refer to Table 1 of \citet{Placco:2014}, which consists of the cumulative CEMP and CEMP-no fractions produced from the 505 stars below [Fe/H] $ <-2.0$ with high-resolution spectroscopy.

Figure~\ref{fig:CEMP_frac} compares the cumulative CEMP fraction of our K-dwarf sample with those from \citet{Yoon:2018:AEGIS} and \citet{Placco:2014}. For the K-dwarf CEMP and CEMP-no fractions, the shaded area represents the uncertainties estimated from the binomial proportion confidence interval \citep[see][for details]{Wilson:1927}. In the left panel of Figure~\ref{fig:CEMP_frac}, we see that the S-PLUS K-dwarf cumulative CEMP fraction is fairly consistent with the estimates from \citet{Yoon:2018:AEGIS}. While the K-dwarf CEMP fraction at [Fe/H] $\le -1.5$ of $20$\,\% is in excess of the \citet{Yoon:2018:AEGIS} estimate of 14\,\%, both studies indicate a CEMP fraction above $60$\,\% at metallicities below [Fe/H] $ = -3.0$. In the right panel of Figure~\ref{fig:CEMP_frac}, we see a larger contrast between the CEMP fraction from the S-PLUS K-dwarf sample with the \citet{Placco:2014} study. While the CEMP fraction at [Fe/H] $\le -2.0$ of the S-PLUS K-dwarf sample, $\sim 27$\,\%, is consistent with the \citet{Placco:2014} value of 33\,\%, the S-PLUS K-dwarf sample rises faster at lower metallicities, somewhat exceeding the \citet{Placco:2014} estimate at [Fe/H]$\le -3.0$.

The sub-class of CEMP stars that exhibit no over-abundances of neutron-capture elements, the CEMP-no stars \citep{Beers:2005}, are of particular interest.  The carbon enhancements in CEMP-no stars are thought to arise from \textit{intrinsic} processes \citep{Yoon:2016}, such as pollution of their natal gas by massive stars, rather than from \textit{extrinsic}  mass-transfer events from an evolved companion in binary systems \citep{Hansen:2016c}. These stars are therefore believed to be true second-generation stars, with chemical characteristics reflecting the elemental signatures produced by first-generation massive stars. While confirmation of CEMP-no stars conventionally requires determination of the barium abundance, [Ba/Fe], \citet{Yoon:2016} demonstrated that CEMP-no stars can be confidently discriminated from CEMP-s and CEMP-r/s stars by their locations on the Yoon-Beers diagram. Following the prescription from \citet{Lee:2017} and \citet{Placco:2018}, we select stars with $A\textrm{(C)}<7.5$, and consider the resulting cumulative CEMP-no fraction in Figure~\ref{fig:CEMP_frac}. 

While qualitatively similar, the crude estimate of the cumulative CEMP-no fraction for the S-PLUS K-dwarf sample consistently exceeds the medium-resolution spectroscopic result from \citet{Yoon:2018:AEGIS} by a small amount, $\sim 5$\,\% at metallicities below [Fe/H$<-1.5$. The difference is even larger when compared with the high-resolution spectroscopic estimate from \citet{Placco:2014}, by as much as $\sim 20$\,\%. However, both the S-PLUS K-dwarf sample and the \citet{Placco:2014} result indicate equivalent cumulative CEMP and CEMP-no fractions at [Fe/H] $<-3.0$, suggesting that the great majority of CEMP stars below [Fe/H]$ = -3.0$ belong to the CEMP-no class. 

The cumulative CEMP fraction produced in this work is generally in best agreement with the medium-resolution result from \citet{Yoon:2018:AEGIS}, supporting the conclusion that photometric techniques such as those discussed in this work are a valid means of probing the CEMP fraction of MW stars. As explored in \citet{Yoon:2020}, the strength of carbon molecular bands in the visible spectrum can become highly disruptive to low- and medium-resolution spectroscopic studies, where determination of the proper pseudo-continuum is both crucial and difficult. This has the effect of limiting the range of CEMP stars that can be studied at these resolutions by means of the CH $G$-band, effectively to $A(\rm{C}) < 8.5$ \citep{Yoon:2020}. Stars in excess of this $A$(C) value are therefore excluded from estimates of the CEMP fraction. It is not clear what effect this has on the frequencies reported from such studies. While estimates of $A$(C) produced with high-resolution spectroscopy are largely immune to the difficulties imposed on pseudo-continuum determination, studies like \citet{Placco:2014} are nevertheless challenged by the significantly smaller numbers of stars with available high-resolution spectra. Interestingly, the number of stars in the S-PLUS K-dwarf sample with [Fe/H] $<-2.0$, 513, is about the same as the 502 stars from \citep{Placco:2014}, which represent nearly all VMP stars with high-resolution carbon measurements available at the time of publication. It is clear that future data releases from S-PLUS will provide substantially larger samples of metal-poor stars with which to further constrain the CEMP fraction in the MW.\\

\subsection{Catastrophic Failures of Photometric Metallicity Estimates}

Common to all techniques of photometric-metallicity estimation is the extreme sensitivity to errors in the adopted metallicity indicator, whether a  filter magnitude or colors, in particular at very low metallicities or high temperatures, where the metallicity dependence of the photometry is diminished. This sensitivity introduces the possibility of ``catastrophic failure", which we define as gross errors by the adopted metallicity estimator(s) of $|\Delta\textrm{[Fe/H]}| > 1.0$\,dex. While the neural-network approach employed in this work is not immune to the occurrence of catastrophic failure, a number of measures have been taken to mitigate this problem. The concensus ANN procedure we have adopted considered the ensemble of networks contributing to the final estimates, weighing each by its validation score achieved during training. Each network is trained on a unique combination of photometric colors, minimizing the influence of any single color on the final estimate. Finally, each network performs an interpolation check during estimation, flagging estimates from inputs outside the parameter space seen during training. This, along with the use of multiple metallicity indicators, results in a very low rate of catastrophic errors for our photometric metallicity estimates, when compared with external spectroscopic estimates: for metallicities below [Fe/H]$<-2.0$, $\sim 0.4$\% for $| \Delta \textrm{[Fe/H]} | > 1.0$\,dex, and only  6\% for $| \Delta \textrm{[Fe/H]} |  > 0.50$\,dex. 
 

The \textit{Pristine} survey has proven quite successful with respect to candidate metal-poor star detection, with success rates for the identification of bona-fide [Fe/H] $<-2.5$ stars of 56\% \citep{Aguado:2019}. The \textit{Pristine} survey exhibits catastrophic failure rates of 1.3\% for $| \Delta \textrm{[Fe/H]} | > 1.0$\,dex, and 9.3\% for $| \Delta \textrm{[Fe/H]} | > 0.50$\,dex for stars with [Fe/H] $<-2.0$, when compared with an overlapping sample of SDSS and SDSS/SEGUE stars \citep[calculations based on data from][]{Starkenburg:2017}.
We emphasize the {\it substantial decrease in the 
catastrophic-failure rates} seen in this study to highlight the advantage of narrow-band photometry (in particular with multiple metallicity-sensitive filters), in conjunction with the sophistication of machine-learning techniques, in this case \texttt{SPHINX}, compared with traditional (often simple polynomial-based) photometric techniques used by many previous efforts. Given the quite low catastrophic-failure rates that are achievable, our photometric approach would appear capable of precision studies of the MDFs of stellar populations without the need for time-consuming medium- or high-resolution spectroscopic follow-up, except perhaps at the very lowest metallicities, e.g., [Fe/H] $\leq -3.0$.

\section{Conclusions}\label{section:conclusions}

Photometric estimates of effective temperature, $T_{\rm eff}$, metallicity, [Fe/H], carbonicity, [C/Fe], and absolute carbon abundances, $A$(C), are produced using the  novel artificial neural network methodology \texttt{SPHINX}, taking advantage of the mixed-bandwidth photometry obtained by the Southern Photometric Local Universe Survey (S-PLUS). Two selections are made from the Stripe 82 data release from S-PLUS, a main-sequence turnoff (MSTO) sample of the inner/outer halo and thick-disk populations ($N=771$), and a main-sequence K-dwarf sample ($N=52,035$), selected based on stellar mass. We study the MDFs for both samples, and conduct an additional exploration of the $A$(C) distribution of our mass-selected K-dwarf sample. 

We reproduce the MDF of the halo with our MSTO sample, with peaks in the distribution corresponding to previously understood populations.  From the mass-selected K-dwarf sample, we produce the first photometric $A$(C) vs. [Fe/H] diagram of the Solar Neighborhood ($1 < d\textrm{(kpc)} < 6$). We constrain the exponential slope of the local-volume MDF, produced with K-dwarf stars, but find significant variation in the value, $\lambda_{10, \textrm{[Fe/H]}} = 0.9\pm0.2$, across the metallicity range considered in this work (which is well-constrained down to $\textrm{[Fe/H]}=-2.75$). Our results suggest that the Solar Neighborhood MDF is better represented by a gamma distribution.

We identify 364 CEMP candidates for follow-up spectroscopic analyses, as well as provide compelling evidence supporting the diverse CEMP morphology discussed in \citet{Yoon:2016}. We find fractions of carbon enhancements similar to previous spectroscopic studies, highlighting S-PLUS as a useful probe of the characteristics of CEMP stars in the MW. Our results confirm that as many as 60\% of extremely metal-poor stars (EMP; [Fe/H] $<-3.0$) are CEMP-no stars. 


While photometric surveys are an ideal means with which to identify candidate metal-poor stars for high-resolution spectrscopic follow-up, the typical magnitude ranges of such surveys (e.g., SDSS with $14.0 < g < 22$) present challenges, as faint targets ($g > 15$) require long integration times with even 6.5-10 meter class telescopes \citep{Mendes:2019}. For this reason, we anticipate the component of S-PLUS known as the Ultra-Short Survey (USS), that will cover the same footprint as the main survey of S-PLUS with exposure times that are 1/12th of main survey. The USS will enable identification of metal-poor candidates as bright as 8th\,mag in the broad-band filters, compared with 12th\,mag for the main survey \citep[see][for details]{Mendes:2019}. In addition to the effectiveness demonstrated in this work of stellar chemical-abundance determinations with S-PLUS, the USS thereby positions S-PLUS as a crucial component for the future of bright metal-poor star studies in the Solar Neighborhood.

The constraints on the metallicity distribution functions of the inner-halo region and Solar Neighborhood produced in this work present a crucial means with which to constrain cosmological models of Milky Way formation and chemical evolution, while circumventing many of the known biases inherent to spectroscopic methods. In particular, we demonstrate the potential to discern the complementary roles of accretion and \textit{in situ} star formation in the chemical evolution of the Milky Way. We expect that robust photometric carbon abundance determinations will play an increasingly important role in the differentiation between cosmological models of Galactic chemical evolution, particularly by more tightly constraining the CEMP and CEMP-no fractions of the inner/outer halo regions and the Solar Neighborhood.

Stellar chemical-abundance determinations with narrow-band photometric surveys like J-PLUS and S-PLUS are a natural application of machine learning algorithms like \texttt{SPHINX}, described in \citet{Whitten:2019a} and implemented in this work. \texttt{SPHINX} has been shown to produce robust estimates of effective temperature, metallicity, carbonicity, and absolute carbon abundance in stars, with a remarkably low rate of catastrophic failure. We anticipate the utility of \texttt{SPHINX} (or similar approaches) for future data releases from J-PLUS and S-PLUS, in particular because several of the narrow-band filters in these efforts will soon be calibrated to enable estimation of additional valuable elemental abundances, such as the $\alpha$-elements Mg and Ca.

\acknowledgments

\textcolor{white}{-}

The authors would like to thank the anonymous referee for providing insightful comments on the manuscript. 
D.D.W., V.M.P., T.C.B., J.Y., and S.R. acknowledge partial support for this work from grant PHY 14-30152; Physics Frontier Center/JINA Center for the Evolution of the Elements (JINA-CEE), awarded by the US National Science Foundation. 
The work of V.M.P. is supported by NOIRLab, which is managed by the
Association of Universities for Research in Astronomy (AURA) under a
cooperative agreement with the National Science Foundation.
D.A. acknowledges support provided by the National Research Foundation of Korea (NRF-2018R1D1A1A02085433). 
Y.S.L. acknowledges support from the National Research Foundation
(NRF) of Korea grant funded by the Ministry of Science and ICT (NRF-2018R1A2B6003961)
F.A-F, F.R.H, and C.E.B. acknowledge funding for this work from FAPESP grants 2019/26492-3, 2019/11910-4, 2018/20977-2, 2018/21661-9, 2016/12331-0 and 2009/54202-8.
H.D.P. acknowledge funding for this work from FAPESP grant 2018/21250-9.
P.B.T. acknowledges partial support from FONDECYT Regular 1200703.
K.Y. gratefully acknowledges funding by the Emmy Noether program from the Deutsche Forschungsgemeinschaft (DFG) and from the European Research Council (ERC) under the European Union’s Horizon 2020 research and innovation programme (Grant agreement No. 852977)
The S-PLUS project, including the T80-South robotic telescope and the S-PLUS scientific survey, was founded as a partnership between the Funda\c{c}\~{a}o de Amparo \`{a} Pesquisa do Estado de S\~{a}o Paulo (FAPESP), the Observat\'{o}rio Nacional (ON), the Federal University of Sergipe (UFS), and the Federal University of Santa Catarina (UFSC), with important financial and practical contributions from other collaborating institutes in Brazil, Chile (Universidad de La Serena), and Spain (Centro de Estudios de F\'{\i}sica del Cosmos de Arag\'{o}n, CEFCA). We further acknowledge financial support from the São Paulo Research Foundation (FAPESP), the Brazilian National Research Council (CNPq), the Coordination for the Improvement of Higher Education Personnel (CAPES), the Carlos Chagas Filho Rio de Janeiro State Research Foundation (FAPERJ), and the Brazilian Innovation Agency (FINEP).
The authors who are members of the S-PLUS collaboration are grateful for the contributions from CTIO staff in helping in the construction, commissioning and maintenance of the T80-South telescope and camera. We are also indebted to Rene Laporte and INPE, as well as Keith Taylor, for their important contributions to the project. 
From CEFCA, we thank Antonio Mar\'{i}n-Franch for his invaluable contributions in the early phases of the project, David Crist{\'o}bal-Hornillos and his team for their help with the installation of the data reduction package \textsc{jype} version 0.9.9, C\'{e}sar \'{I}\~{n}iguez for providing 2D measurements of the filter transmissions, and all other staff members for their support with various aspects of the project.

\clearpage
\bibliographystyle{aasjournal}
\nocite{*}
\bibliography{main.bib}

\section{Appendix}

\begin{figure*}[h]
	\centering
	\includegraphics[width=\textwidth, trim=1.80cm 2.50cm 2.0cm 2.0cm, clip]{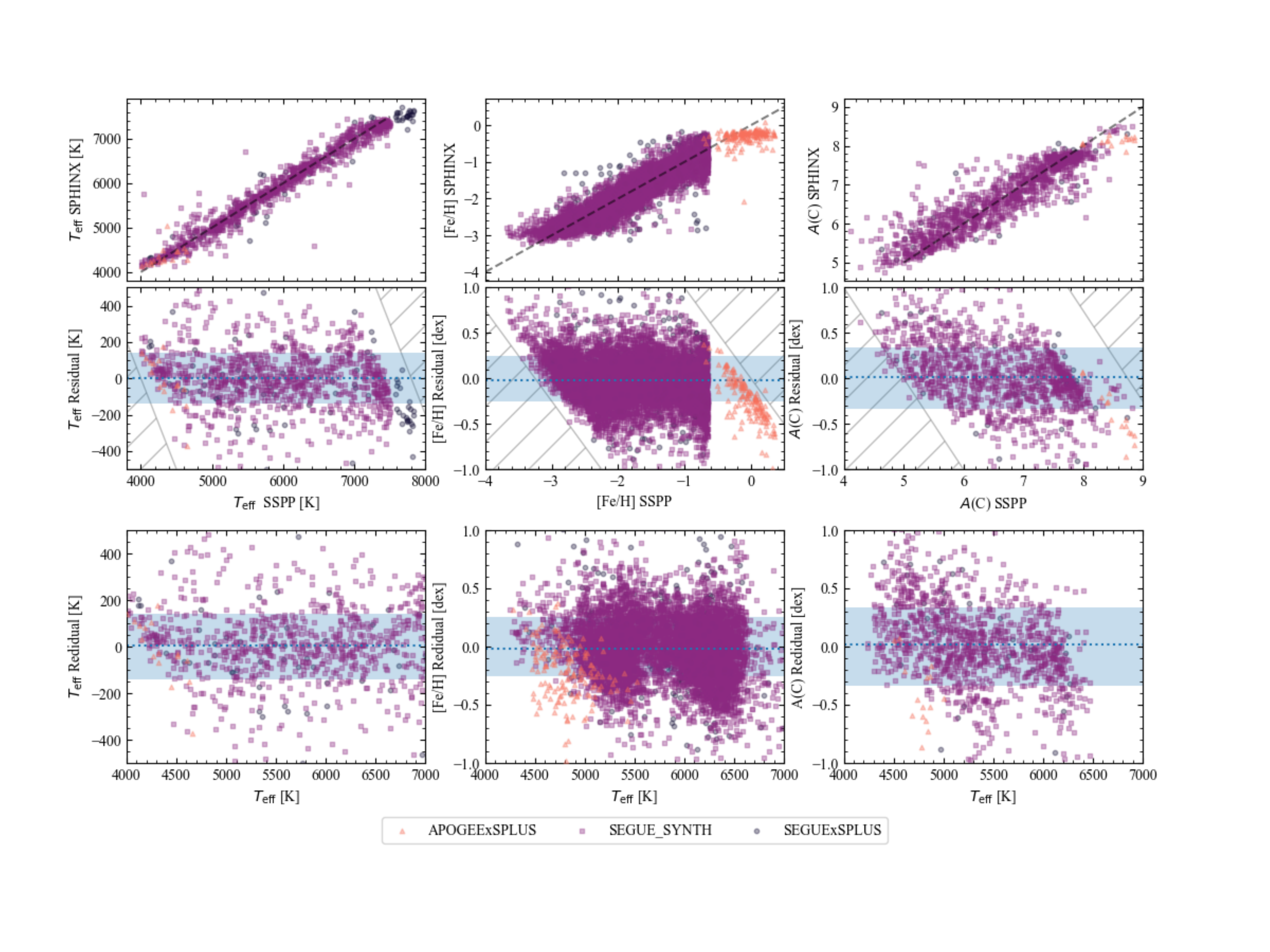}
	
	\caption{Photometric estimates of effective temperature are shown in the left panels, with metallcity and absolute carbon abundance in the center and right panels, respectively. All results are produced by \texttt{SPHINX} using S-PLUS photometry. In the bottom panels, residuals are shown of effective temperature, metallicity, and absolute carbon abundance, as a function of the spectroscopic effective temperature estimate from the n-SSPP estimates of SEGUE and APOGEE  stars.
	\label{fig:residual_parameters}}
\end{figure*}

In this section, we provide figures relevant to the calibrations for \texttt{SPHINX}. In Figure~\ref{fig:residual_parameters}, the residuals for effective temperature, metallicity, and absolute carbon abundance are shown, determined with an independent validation set during the training procedures for \texttt{SPHINX}, as described in the text.

\end{document}